\documentclass[10pt,twocolumn]{IEEEtran}
\makeatletter
\def\ps@headings{%
\def\@oddhead{\mbox{}\scriptsize\rightmark \hfil \thepage}%
\def\@evenhead{\scriptsize\thepage \hfil \leftmark\mbox{}}%
\def\@oddfoot{}%
\def\@evenfoot{}}
\makeatother
\usepackage{amssymb}
\usepackage{graphicx}
\usepackage{cite}
\usepackage{bm}

\def\bq{\begin{equation}}
\def\eq{\end{equation}}
\def\bqn{\begin{eqnarray}}
\def\eqn{\end{eqnarray}}
\def\bqnn{\begin{eqnarray*}}
\def\eqnn{\end{eqnarray*}}

\def\bfx{{\bf x}}

\def\bfone{{\bf 1}}
\def\defeq{\stackrel{\mathrm{def}}{=}}

\newcommand{\sizex}[1]{| #1 |_2}
\newcommand{\lengthx}[1]{| #1 |_1}
\newcommand{\mod}[2]{\langle #1,#2 \rangle}

\newcommand{\senseResult}[3]{{#1}_d^{(#2)}(#3)}
\title{Estimating Shape of Target Object Moving on Unknown Trajectory by Using Location-Unknown Distance Sensors: Theoretical Framework}
\author{Hiroshi~Saito,~\IEEEmembership{Fellow,~IEEE} and Hiroki Ikeuchi
\thanks{Manuscript received }
\thanks{Hiroki Ikeuchi is with NTT Network Technology Laboratories, 3-9-11, Midori-cho, Musashino-shi, Tokyo 180-8585, Japan, E-mail: ikeuchi.hiroki@lab.ntt.co.jp}
}

\date{}

\begin{document}

\maketitle
\begin{abstract}
By using directional distance sensors that have unknown locations, this paper proposes a method of estimating the shape of a location-unknown target object $T$ moving with unknown speed on an unknown straight line trajectory.
Regardless of many unknown factors, the proposed method can estimate the shape by using each sensor's continuous report of the measured distance to $T$ without using side information or additional mechanisms such as locations of anchor sensors and angle-of-arrival measurements.
By using the sensor reports, the proposed method estimates (i) the moving speed of $T$, (ii) the length and direction of an edge of $T$, and (iii) the order of consecutive edges.
As a result, we can obtain the shape of $T$.

\end{abstract}

\section{Introduction}
A few decade has passed since the proposal of a new sensing paradigm using a small and low-cost sensors like dust \cite{smartdust}.
That is, instead of having a few sensors with advanced functions and high performance, this paradigm has many sensors with simple functions and low performance \cite{wins,survey,sensornode,survey2}.
They are networked through a wireless link and send reports, each of which includes only an insignificant amount of information but may give us significant information if we collect all.
Because many sensors are deployed in this new sensing paradigm, we cannot carefully plan the location of each one. 
A global positioning system (GPS) cannot be used because the sensors should have limited capability and keep power consumption low.
Developing low power wide area networks (LPWANs), such as LoRa WAN \cite{lora}, narrow-band Internet of Things (NB IoT) \cite{nb_iot}, a wide area ubiquitous network \cite{commag,VTC}, and SIGFOX \cite{sigfox}, supports sensors with low performance and functionalities and a long-range, low-speed wireless link with very low power consumption.
These networks will enable us to implement this sensing paradigm and IoT \cite{IoT,IoTsurvey}.

For a challenging application under this new sensing paradigm, we investigated the problem of estimating target-object shape by using randomly distributed distance sensors.
A distance sensor is often composed of a pair of transmitter/emitter such as an infrared emitting diode, an ultrasound, a laser and a detector detecting its reflection.
We can find a commercial sensor of a few US\$ that is palm-sized or smaller. 
Such sensors may be deployed for various applications such as security surveillance.
The original objective of deployment of such sensors may not be to estimate the shape of a target object.
However, this paper demonstrates that simple distance sensors deployed randomly at unknown locations can be applied to estimate a target-object shape at an unknown location and moving speed.
If each sensor is provided by an independent third party, this suggests that crowdsensing (participatory sensing) using directional distance sensors can enable us to estimate the target-object shape while maintaining location-privacy.

An individual sensor in this paper is a simple sensor measuring the distance between a sensor and a target object and has communication capability.
It does not have a positioning function, such as a GPS, and it is placed without careful design.
By collecting reports from individual sensors, we statistically estimate the target's object shape and moving speed.

The contributions of this paper are:

\begin{itemize}
\item
This paper demonstrates that we can estimate the shape of a moving target object with location-unknown distance sensors in an unknown sensing direction.
Those sensors continuously sense distance and report the sensing result.
The estimation method does not need any positioning function, anchor location information, or additional mechanisms to obtain side information such as angle of arrival of signal. 
Although previous studies suggest that shape estimation is impossible with location-unknown simple sensors, this paper shows that continuous sensing enables us to estimate the shape of the target object moving on an unknown line by using simple sensors with unknown locations.
For the time-invariant polygon target object, the proposed estimation method estimates each edge's length and direction and the connectivities of edges to estimate the complete shape of the target object.
To my best knowledge, this is the first paper proposing the shape estimation method under such conditions.

\item
A moving speed of the location unknown target object is also estimated with the location unknown distance sensors.
\end{itemize}

In addition to its explicit contributions, this paper suggests that, because location information of sensors is not essential to estimate a target-object shape, various estimations using location-unknown sensors may be possible.
This is important for crowdsensing or participatory sensing from the location-privacy point of view.

\section{Related works}
The fundamental questions for this problem under this paradigm are whether we can estimate the shape of a target object under the new paradigm using many small simple sensors and how we estimate if possible.
Previous studies suggested that only a small number of parameters such as the size and perimeter length of a target object can be estimated with randomly deployed unknown-location simple sensors such as binary sensors and distance sensors and other parameters cannot be estimated \cite{infocom,ieice-invite,arXiv}.
Thus, composite sensors, which are composed of several simple sensors, were introduced.
They are randomly deployed and their locations are unknown \cite{signalProcess,mobileComp}.
By using them, additional parameters can be estimated.
Unfortunately, however, they are difficult to implement and deploy, particularly when the composite sensors are large.
Those studies used the sensing results at a certain sensing epoch, and estimated parameters using them.
Even when they used the sensing results at multiple sensing epochs, they did not take account of sensing epoch information.
Only one \cite{time-variant} in that series took account of sensing epochs and the temporary behavior of sensing results, but it focused on estimating the size and perimeter length of the target object.

A new study has recently estimated the shape of a fixed target-object by using mobile distance sensors with unknown locations \cite{new}.
The estimation method structure in which parts of the target object and their connectivities are estimated is similar, but there are major differences between this paper and that paper.
(i) The estimation in this paper needs to estimate the target object's moving speed, but that in the other paper assumes known sensor moving speeds.
(ii) A single mobile sensor's report enables the relative edge direction (the edge direction from the moving direction of the sensor) and the edge length to be estimated, but the estimation in this paper does not.
This is mainly because the sensing area direction is unknown in this paper.

To the best of my knowledge, no studies other than the above have directly tackled these questions.
However, there have been considerable amount of studies on developing an estimation method using location-unknown sensors.
These studies took a different approach.
Most first estimated the sensor locations \cite{locating_nodes} because it is believed that ^^ ^^ the information gathered by such sensor nodes, in general, will be useless without determining the locations of these nodes" \cite{flip_amb} or ^^ ^^ the measurement data are meaningless without knowing the location from where the data are obtained" \cite{local_4}.
Once sensors' locations are estimated, shape estimation is no longer difficult.
However, the approach of estimating the sensor locations often requires additional mechanisms or side information, such as locations of anchor sensors and measurement mechanisms including angle-of-arrival measurements, training data and period, and distance-related measurements \cite{locating_nodes,local_2,local_3,local_4}.
Concrete examples are intersensor distance information \cite{flip_amb}, location-known anchor sensors \cite{tsp2002}, set of signals between sensors \cite{acm_sensor}, and the system dynamic model and location ambiguity of a small range \cite{bernoulli}.

In addition, there has been research capturing the shape of a target object by using cameras that cannot cover the whole shape of the target object \cite{camera}.

\section{Model}
The target object $T$ is coming into and going out of a monitored area $\Omega\subset\mathbb{R}^2$.
It is moving at an unknown constant speed $v>0$ along an unknown reference directional line.
In the remainder of this paper, we use this directional line as the $x$-axis and its direction as the reference direction.
(We do not need to know the reference direction.  This is just used to define direction.)
$T(t)\subset\mathbb{R}^2$ denotes the set occupied by $T$ at $t$.

\begin{figure}[tb] 
\begin{center} 
\includegraphics[width=8cm,clip]{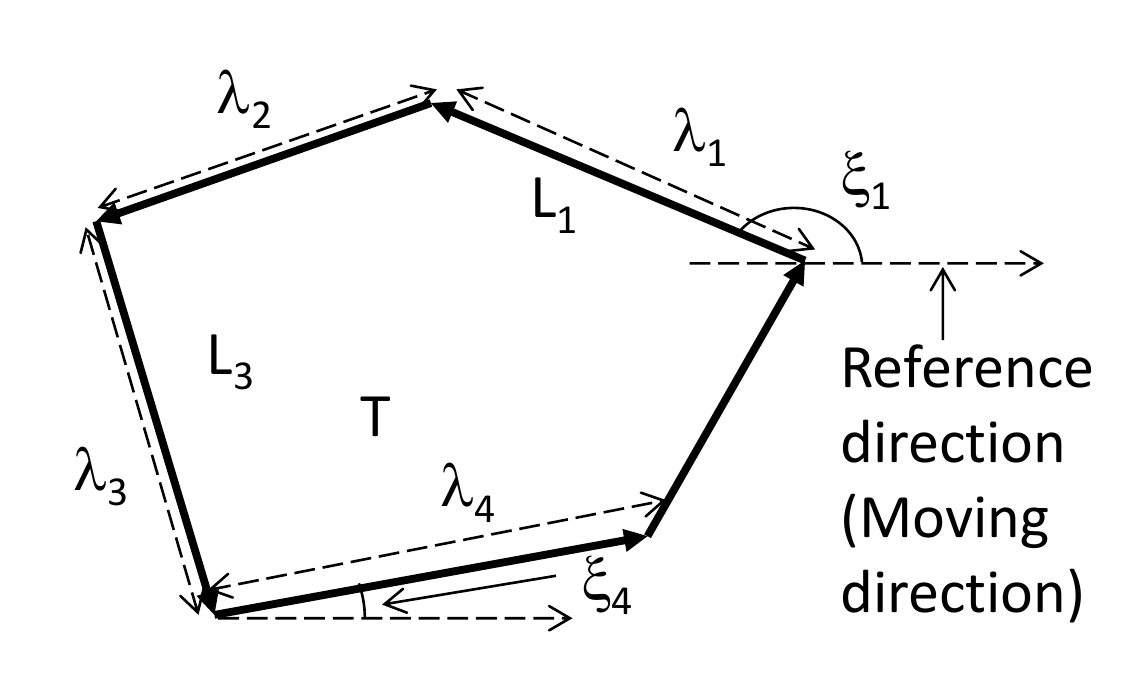} 
\caption{Illustration of target object model} 
\label{model} 
\end{center} 
\end{figure}

$T$ is a polygon, and its boundary $\partial T$ is closed and simple (no holes or double points) and consists of directional edges $\{L_j\}_j$ where $j \geq 1$ (Fig. \ref{model}).
Let $\lambda_{j}$ be the length of $L_{j}$, and let $\xi_{j}$ be the angle formed by $L_{j}$ and the reference direction where $0\leq \xi_j<2\pi$.
Here, $\{L_{j}\}_j$ are counted counterclockwise along $\partial T$ and the head of $L_{j}$ is the tail of $L_{j+1}$.
We do not know any of $\{\lambda_{j}, \xi_{j}\}_{j}$.
That is, we do not know the target-object shape, size, or location.

There are $n_s$ directional distance sensors deployed in $\Omega$.
Each sensor can continuously measure the distance to an object lying in the sensing direction within the maximum range $r_{max}>0$. 
Therefore, when the location of a sensor is $\bfx=(x,y)$ and its sensing direction from the reference direction is $\theta$, the sensing area $S(\bfx,\theta)$ is $\{(x+s\cos\theta,y+s\sin\theta), 0\leq \forall s\leq r_{max}\}$ and the measured distance $r(t)$ at $t$ to $T$ by this sensor is given as follows. 
\bq
r(t)=\cases{\min_{(x+s\cos\theta,y+s\sin\theta)\in T(t)} s, &for  $S(\bfx,\theta)\cap T(t)\neq \emptyset$,\cr
\emptyset, &for $S(\bfx,\theta)\cap T(t)=\emptyset$.}
\eq 
In particular, $r(t)=0$ if $(x,y)\in T(t)$.

These sensors are independently and randomly deployed with each other, and their locations are independent of $T$.
Their directions are also random and independently and uniformly distributed in $[0,2\pi)$.
For the $i$-th sensor ($1\leq i\leq {n_s}$), let $\bfx_i$ be its location, $\theta_i$ be its direction, and $r_i(t)$ be the measured distance to $T$ at $t$.
Assume that we do not know $\bfx_i$ or $\theta_i$ for any $i$.
That is, we do not know their locations or directions.
We may remove the subscript and use $L$, $\bfx$, $\theta$, $r(t)$, and $\xi$ to simplify the notation.
Because sensors monitor $\Omega$, assume that $\bfx_i\subset \Omega$ for all $i$.
To remove the boundary effect of $\Omega$, assume $\sizex{\Omega}\gg r_{max}^2,\sizex{T}$ where $\sizex{X}$ is the area size of the set $X\subset \mathbb{R}^2$.

Each sensor can communicate with a server collecting sensing reports from individual sensors.
It reports the measured distance $r(t)$ between the sensor and a target object if it detects within the maximum sensing range or reports ^^ ^^ no detection" otherwise.
Because it does not have a positioning function or direction information of the sensor, the report does not include $\bfx$ or $\theta$.
All the sensors are assumed to continuously send reports.

In the remainder of this paper, we use the following notations.
$\bfone(z)\defeq\cases{1, &if $z$ is true,\cr 0, &otherwise,}$ and $\widehat{z}$ is an estimator of $z$.
For angles $t_1,t_2$, $\mod{t_1}{t_2}$ is an interval $[t_1,t_2)$ under mod $2\pi$.  
That is, $\mod{t_1}{t_2}$ is an interval $[t_1,t_2)$ if $t_1,t_2<2\pi$ and is intervals $[t_1,2\pi)\cup [0,t_2-2\pi)$ if $t_1<2\pi, 2\pi\leq t_2<4\pi$.

Table \ref{p_list} lists the variables and parameters used in the remainder of this paper for the reader's convenience.

\begin{table}
\caption{List of variables and parameters}
\begin{center}\label{p_list}
\begin{tabular}{ll}
\hline
$\Omega$&monitored area\\
-- Target object --\\
$T$&target object\\
$T(t)$&the set occupied by $T$ at $t$ in $\mathbb{R}^2$\\
$L_{j}$&$j$-th directional line segment of $\partial T$\\
$\lambda_{j}$&length of $L_{j}$\\
$\xi_{j}$&angle formed by $L_j$ and reference direction\\
$v$&moving speed of $T$\\
-- Sensor --\\
$n_s$& number of sensors\\
$r_{max}$&maximum sensing range\\
$\bfx_i, \theta_i$&$i$-th sensor's location and direction\\
$r_i(t)$&measured distance to $T$ by $i$-th sensor\\
-- Basic properties --\\
$p_d(L|\theta)$&period detecting whole $L$ by sensor of direction $\theta$\\
$l_d(L|\theta)$&length in time of $p_d(L|\theta)$\\
$s_d(L|\theta)$&slope of $r(t)$ during $p_d(L|\theta)$\\
$n_d$& number of $(l_d,s_d)$ pairs\\
$m_t$&total sensing time while $T$ is $\Omega$\\
$\theta_{min1}(\xi)$&$\theta_{min1}(\xi)\defeq\arcsin(\lambda|\sin\xi|/r_{max})$\\
$\Psi(x)$&set of sensing results satisfying $x$\\
$\tilde{\lambda},\tilde{\xi}$&temporary estimates of $\lambda,\xi$\\
$\widehat{n_e}(\lambda,\xi)$&estimated number of edges with $(\lambda,\xi)$ (Eq. (\ref{def_ne}))\\
$n_c(a,a';b,b')$&number of consecutive sensing results belonging \\
&to $\Psi_{a,a'}$ and $\Psi_{b,b'}$\\
\hline
\end{tabular}
\end{center}
\end{table}

\section{Basic properties}\label{basic_pro}
\begin{figure}[tb] 
\begin{center} 
\includegraphics[width=9cm,clip]{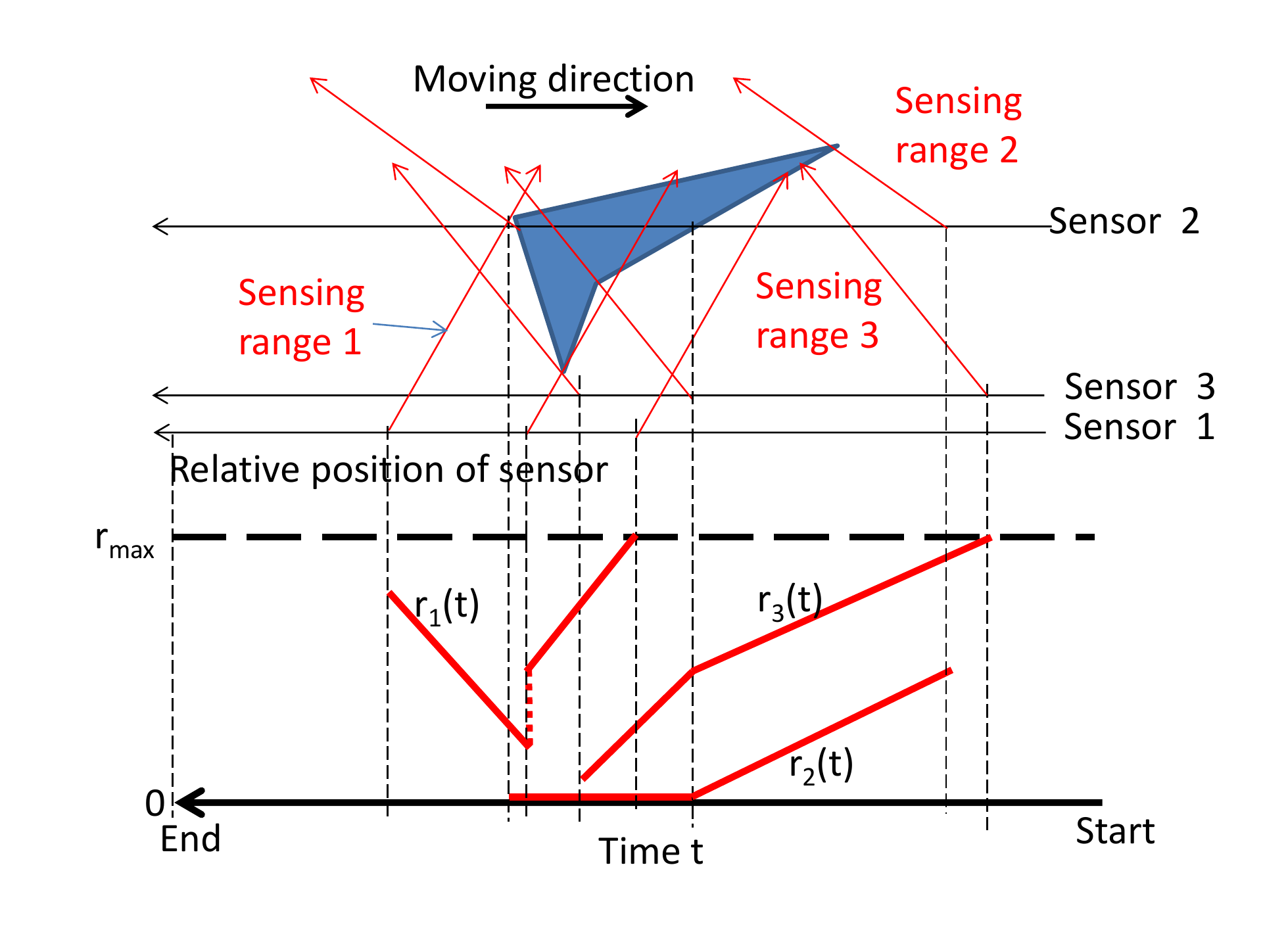} 
\caption{Basic example} 
\label{basic} 
\end{center} 
\end{figure}

This section discusses basic properties of $r(t)$.
Let us see a simple example illustrated in Fig. \ref{basic}.
Instead of moving target objects, the relative position of the sensor is moving in this figure.
The sensor detects only a single edge located the nearest to the sensor and its distance from the sensor is $r(t)$.
Note that $r(t)=0$ occurs if and only if the sensor is in $T(t)$.

An important observation of this figure is that there may be some jumps in $r(t)$ from a certain value between 0 and $r_{max}$ to another certain value ($r_1(t)$ in Fig. \ref{basic}).  
Only a single edge located the nearest to a sensor is detected by the sensor and its distance from the sensor is $r(t)$.
Although other edges are within a sensing area, they are not detected or their distances to the sensor are not measured.
That is, a detection of an edge may be blocked by another edge.
A jump down (up) occurs when a blocked detection starts (finishes).

\begin{figure}[tb] 
\begin{center} 
\includegraphics[width=9cm,clip]{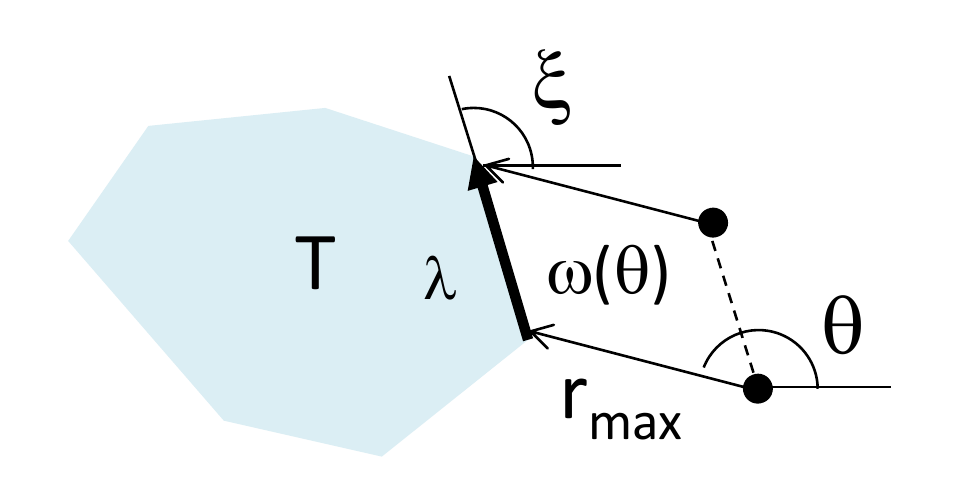} 
\caption{$\omega(\theta)$} 
\label{omega} 
\end{center} 
\end{figure}

A sensor detects an edge $L$ for a given $\theta$ if and only if the sensor is located in $\omega(\theta)$, where $\omega(\theta)$ is a parallelogram attached to the right-hand side of $L$ and one of its edges is $L$ and another edge has the length $r_{max}$ and the direction $\theta$ (Fig \ref{omega}). 
If and only if $\theta\in\mod{\xi}{\xi+\pi}$, $\omega(\theta)$ exists.
That is,
\bq
\theta-\xi\in [0,\pi)\, {\rm mod}\, 2\pi.\label{theta-xi}
\eq

Pay attention to a case in which a sensor keeps detecting $L_{j}$ (Fig. \ref{single-edge}).
When a sensor keeps detecting $L_{j}$, $r(t)$ becomes continuous.
Because $L_{j}$ is a line segment, $r(t)$ becomes a line segment while the sensor keeps detecting it (Fig. \ref{basic}).
When the period $p_d(L_{j}|\theta)$ detecting the whole $L_{j}$ with $r(t)>0$ by a sensor the direction of which is $\theta$ starts at $t_s$ and ends at $t_e$, an event corresponding to $t_s$ is (i) a change of slope at $r(t_s)>0$, (ii) a jump down of $r(t)$ at $t_s$, or (iii) $r(t_s)<r_{max}$ and $r(t_s-dt)=\emptyset$ and an event corresponding to $t_e$ is (i) a change of slope at $r(t_e)>0$, (ii) a jump up of $r(t)$ at $t_e$, or (iii) $r(t_e-dt)<r_{max}$ and $r(t_e)=\emptyset$.
In the remainder of this paper, we use data regarding $p_d(L_{j}|\theta)$ the abovementioned start and end events of which exist if we do not explicitly indicate otherwise.
This suggests that we observe the whole $L_{j}$ with $r(t)>0$ during $p_d(L_{j}|\theta)$.

\begin{figure}[tb] 
\begin{center} 
\caption{Case when sensor keeps detecting $L_{j}$} 
\label{single-edge} 
\end{center} 
\end{figure}

\begin{figure}[tb] 
\begin{center} 
\includegraphics[width=9cm,clip]{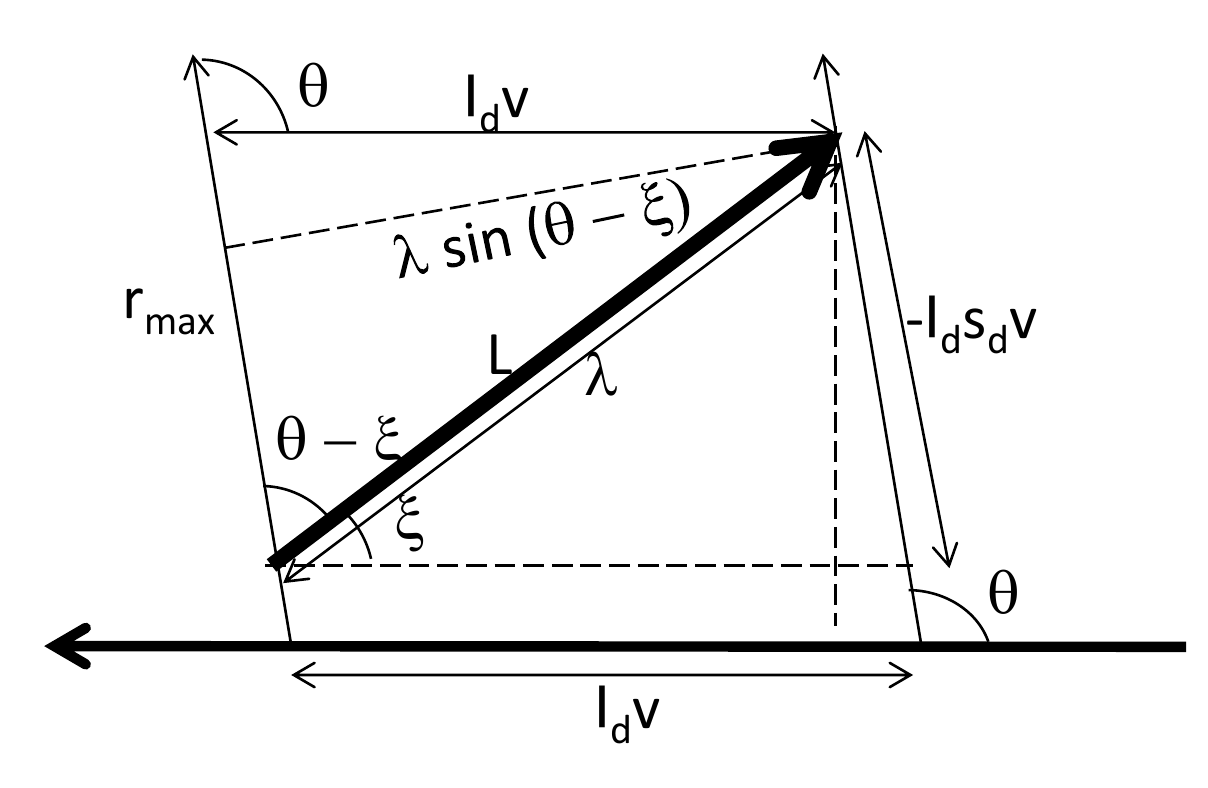} 
\caption{Illustration of $l_d$ and $s_d$} 
\label{ldsdFig} 
\end{center} 
\end{figure}

Let $l_d(L|\theta)$ be the length in time of $p_d(L|\theta)$ where $p_d(L|\theta)$ is the period detecting the whole $L$ with $r(t)>0$ when the sensor direction is $\theta$.
Define  $s_d(L|\theta)\defeq \frac{r(t_e)-r(t_s)}{vl_d(L|\theta)}$ where $p_d(L|\theta)$ starts at $r(t_s)$ and ends at $r(t_e)$.
That is, $s_d(L|\theta)$ is the slope of the $r(t)$ graph during $p_d(L|\theta)$ where the $x$-axis of the graph is the moving length of $T$.
Due to geometric calculation (Fig. \ref{ldsdFig}),
\bqn
vl_d(L|\theta)|\sin\theta|&=&\lambda\sin(\theta-\xi),\label{l_d}\\
-vl_d(L|\theta)s_d(L|\theta)|\sin\theta|&=&\lambda\sin\xi. \label{ldsd}
\eqn
Thus, 
\bq
s_d(L|\theta)=-\sin\xi/\sin(\theta-\xi).\label{s_d}
\eq

The following are the basic properties of $l_d(L|\theta)$ and $s_d(L|\theta)$.

Because of Eq. (\ref{s_d}), 
\bqn
|s_d(L|\theta)|&\geq& |\sin\xi|.\label{abs_sin_xi}
\eqn
Because of Eq. (\ref{ldsd}) and $v,l_d(L|\theta),\lambda,|\sin\theta|\geq 0$, 
\bq
s_d(L|\theta)/\sin\xi\leq 0.\label{sdsinxi}
\eq
Because of Eqs. (\ref{l_d}) and (\ref{s_d}), 
\bq
\{s_d(L|\theta)=0\}\Leftrightarrow\{\xi=0,\pi; \lambda=vl_d(L|\theta)\}.\label{slope0}
\eq

\subsection{For a single pair of $l_d$ and $s_d$}
When $l_d(L|\theta)$ and $s_d(L|\theta)$ are results of detecting an edge $L$ of length $\lambda$ and direction $\xi$, we can describe $\xi$ as a function of $\lambda$.
Here, note that we do not know $\theta$ or which edge we will estimate.

Due to Eq. (\ref{s_d}), $(-1+s_d\cos\theta)\sin\xi=s_d\sin\theta\cos\xi$.
Apply Eq. (\ref{ldsd}) to $\sin\theta$ in this and obtain the following.
\bqn
\cos\xi&=&\pm\mu\cr
\mu&\defeq&\frac{(\lambda/v)^2+l_d^2(1-s_d^2)}{2\lambda l_d/v}\label{xi-by-lam}
\eqn
Because of Eq. (\ref{sdsinxi}), 
\bq
\xi=\cases{
\xi_0, \pi-\xi_0, &if $s_d<0$,\cr
-\xi_0, -\pi+\xi_0, &if $s_d\geq 0$,
}\label{xi-sol}
\eq
 where $\xi_0\defeq \arccos\mu\in [0,\pi)$.

[Remark] Eq. (\ref{xi-sol}) means that we cannot uniquely determine $\xi$.
In fact, we cannot distinguish $T$ from its mirror image only through directional distance sensors randomly deployed.

\subsection{For two pairs of $l_d$ and $s_d$ detecting a single edge}\label{two-sol}
Assume that two sensors detect the same edge of length $\lambda$ and direction $\xi$ and that their sensing results are $(l_d,s_d)$ and $(l_d',s_d')$, respectively.
Because of Eq. (\ref{xi-by-lam}), we obtain $\lambda$:
\bq
\lambda=v\sqrt{\frac{l_dl_d'}{l_d'- l_d}\{ l_d'(1-s_d'^2)-l_d(1-s_d^2)\}}.\label{lam-sol}
\eq
By using $\lambda$ in Eq. (\ref{xi-sol}), we obtain $\xi$.

\subsection{Number of results sensing a whole edge with $r(t)>0$}\label{num_result}
The estimation method proposed in this paper uses sensing result pairs $(l_d,s_d)$ derived from $r(t)>0$.
Here, the expectation of the number $n_d$ of such sensing result pairs is derived.
This is used to estimate the number of edges.

\begin{figure}[thb] 
\begin{center} 
\includegraphics[width=9cm,clip]{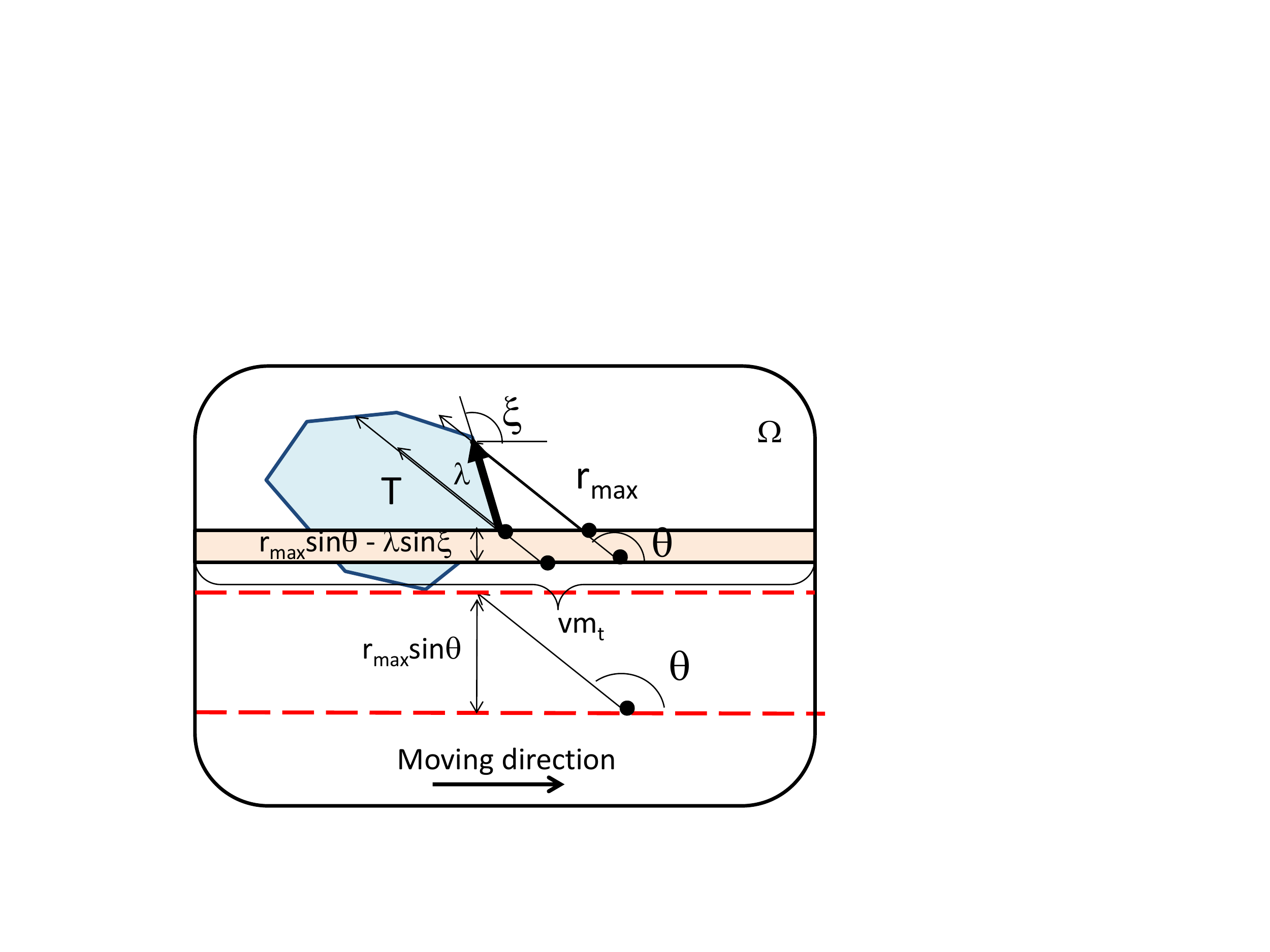} 
\caption{Location of sensors detecting whole edge with $r(t)>0$} 
\label{omega2} 
\end{center} 
\end{figure}

Consider an edge the length of which is $\lambda$ and direction of which is $\xi$.
A sensor the sensing direction of which is $\theta$ detects this whole edge with $r(t)>0$, if $\theta\in\mod{\xi}{\xi+\pi}$ and this sensor is located in a strip the width of which is $r_{max}|\sin\theta|-\lambda|\sin\xi|$ (Fig. \ref{omega2}).
Because the width of this strip must be positive, 
\bq
\theta\in[\theta_{min1},\pi-\theta_{min1})\cup[\pi+\theta_{min1},2\pi-\theta_{min1}),
\eq
where $\theta_{min1}(\xi)\defeq\arcsin(\lambda|\sin\xi|/r_{max})$ for $\lambda|\sin\xi|\leq r_{max}$.
(Note $\theta_{min1}\in[0,\pi/2]$.)
Because the strip length inside $\Omega$ is $vm_t$ and the sensor density is $n_s/\sizex{\Omega}$, $E[n_d(\lambda,\xi)]$ for $\lambda|\sin\xi|\leq r_{max}$ is given as follows where $m_t$ is the total sensing time start from the epoch $T$'s entering $\Omega$ and to the epoch $T$'s leaving from $\Omega$.
\bqn
&&E[n_d(\lambda,\xi)]\cr
&=&vm_tn_s\frac{\int_{\theta\in\Theta_0}(r_{max}|\sin\theta|-\lambda|\sin\xi|) d\theta}{2\pi\sizex{\Omega}}\cr
&=&vm_tn_s\frac{2r_{max}\cos\theta_{min1}-(\pi-2\theta_{min1})\lambda|\sin\xi|}{2\pi\sizex{\Omega}},\label{detect_num}\cr
&&
\eqn
where $\Theta_0\defeq ([\theta_{min1},\pi-\theta_{min1})\cup[\pi+\theta_{min1},2\pi-\theta_{min1}))\cap[\xi,\xi+\pi)$.
(The start epoch is detected by the first epoch a sensor detects $T$ and the end epoch detected by the epoch none of the sensors detect $T$.)
For $\lambda|\sin\xi|> r_{max}$, $E[n_d(\lambda,\xi)]=0$.

\section{Target-object shape estimation}
Now, we are in a position to discuss target-object shape estimation.
Additionally, we estimate $v$.
As a preliminary, we need to obtain $(l_d,s_d)$ from the measured distance data $r(t)>0$.

The shape estimation method consists of five parts.
The first part estimates the target object speed $v$.
Because sensing results depend on $v$, estimating $v$ is an important first step for estimating the target-object shape.
The second step estimates the edges parallel to the $x$-axis, which is the moving direction of $T$.
Because there are many examples of edges of target objects being along the moving direction, that is, $\xi=0,\pi$ and because the estimation becomes very easy for $\xi=0,\pi$, it is worth treating the edges parallel to the $x$-axis as a special case.
The third part estimates the lengths and directions of the other edges.
The second and third parts implicitly include the estimation of the number of edges of $T$.
The fourth part estimates of the order of the edge.
That is, it determines a consecutive edge of a certain edge.
The first to fourth parts should provide the shape of the target object $T$, but we typically may not find the complete shape of $T$ when we fail to estimate an edge.
Particularly when $T$ is not convex, it is likely that we fail to estimate edges forming concave parts of $T$.
The fifth part makes up for errors for estimating edges forming concave parts of $\partial T$.

Because sensors are randomly distributed over $\Omega$, we can obtain enough sensors that have sensing results for a whole edge if $r_{max}$ is sufficiently long and $n_s$ is sufficiently large.
Assume that we obtain $\{(\senseResult l j i, \senseResult s j i)\}_j$ from the sensed distance $r_i(t)$ of the $i$-th sensor where $\senseResult l j i$ ($\senseResult s j i$) is the $j$-th $l_d$ ($s_d$) derived as its sensing result.
Here, note that we do not know which edge we obtain $(\senseResult l j i, \senseResult s j i)$.
This makes the estimation problem difficult and unique.

\subsection{First part: estimating moving speed of $T$}
Obtain the number of sensors detecting $T$ with $r(t)\neq 0$ for any $t$, and derive its expectation as a function of $v$ to estimate $v$.
Note that such sensors are in the red dotted-line strip in Fig. \ref{omega2} the width of which is $r_{max}\sin\theta$ for a given $0\leq \theta<\pi$.
(For a given $\pi\leq \theta<2\pi$, another strip just above $T$ the width of which is $r_{max}|\sin\theta|$.)
Because sensor density is $n_s/\sizex{\Omega}$ and its strip size is $vm_tr_{max}\sin\theta$,
\bqn
&&E[\sum_{i=1}^{n_s} \bfone(r_i(\exists t)>0,r_i(\forall t)\neq 0)]\\
&=&2vm_tn_sr_{max}\int_0^\pi\sin\theta d\theta/(2\pi\sizex{\Omega})\\
&=&2vm_tn_sr_{max}/(\pi\sizex{\Omega}).
\eqn
Let $n_r$ be the measured sample of this number.
Then,
\bq
\widehat{v}=\pi n_r\sizex{\Omega}/(2m_tn_sr_{max}).
\eq

[Remark]
$vm_t$ is used as the length of a strip in $\Omega$ in the first part to estimate $v$ and the second and third parts to determine the number of edges.
The measured $vm_t$ is not exactly the length of a strip in $\Omega$ and can deteriorate the accuracy if $vm_t\gg \max(r_{max}, \lengthx{T})$ is not satisfied where $\lengthx{X}$ is the perimeter length of $X\subset\mathbb{R}^2$.
If $\lengthx{\Omega}\gg r_{max}$ is not satisfied, the assumption of this strip makes the estimate less accurate.

\subsection{Second part: estimating edges parallel to moving direction of $T$}\label{part0direction}
Let $\Psi(s_d=0)$ be the set of samples (sensing results) $\{(\senseResult l j i, \senseResult s j i)\}_{i,j}$ satisfying $\senseResult s j i=0$.
Due to Eq. (\ref{slope0}), $\senseResult s j i =0$ means
\bq
\xi=0,\pi; \lambda=vl_d.
\eq
Therefore, these are basic estimates of $\lambda$ and $\xi$.
There can be multiple edges parallel to the moving direction of $T$.
$\Psi(s_d=0)$ corresponds to multiple edges of different lengths with directions 0 or $\pi$.
Otherwise, $\Psi(s_d=0)$ corresponds to multiple edges of the same length or a single edge.

When we have no idea how many edges are parallel to the moving direction, it is a good idea to apply a classification tool such as Mclust of R \cite{mclust} to the set of $\{\senseResult l j i\}_{i,j}$ in $\Psi(s_d=0)$.
Such a classification tool can divide $\Psi(s_d=0)$ into several subsets $\{\Psi_{0,k}(s_d=0)\}_k$.
For each subset, obtain estimates.
That is, obtain the edge length estimate $\widehat{\lambda}(\Psi_{0,k}(s_d=0))$ by computing the mean of $\widehat{v}l_d, l_d\in \Psi_{0,k}(s_d=0)$.
Note that $\widehat{\xi}(\Psi_{0,k}(s_d=0))=0,\pi$ for any $k$.

The number of edges to which $\Psi(s_d=0)$ or its subset corresponds can be estimated through $E[n_d(\lambda,\xi)]$ given by Eq. (\ref{detect_num}) when the edge length $\lambda$ and direction $\xi$ are given. 
For given $\widehat{\lambda}$ and $\widehat{\xi}$, $\widehat{n_e}(\widehat{\lambda},\widehat{\xi})$ defined below (approximately) provides the number of edges corresponding to $\Psi(s_d=0)$ or its subset, where $\tilde n_d(\lambda,\xi)$ is the observed $n_d(\lambda,\xi)$ and 
\bq
\widehat{n_e}(\widehat{\lambda},\widehat{\xi})\defeq\tilde n_d(\widehat{\lambda},\widehat{\xi})/E[n_d(\widehat{\lambda},\widehat{\xi})].\label{def_ne}
\eq
$\widehat{n_e}(\widehat{\lambda},\widehat{\xi})$ enables us to determine the number of edges corresponding to $\Psi(s_d=0)$ or its subset, where $\widehat{\lambda},\widehat{\xi}$ are estimates derived by $\Psi(s_d=0)$ or its subset.
$\tilde n_d(\widehat{\lambda},\widehat{\xi})$ is given by the number of samples in $\Psi_{0,k}(s_d=0)$ such that $\widehat{\lambda}$ is derived by using these samples.


\subsection{Third part: estimating edges in general}
This part consists of (i) temporary estimation of the length and direction of each edge, (ii) evaluation of the number of sensing results consistent with the temporary estimation, (iii) decision of whether the temporary estimation is adopted, and  (iv) estimation of number of edges.
We adopt the temporary estimated length and direction with which many sensing results are consistent as their estimates $\widehat{\lambda},\widehat{\xi}$.
This idea comes from the fact that, if the estimates for $L_j$ are exact, sensing results detecting $L_j$ are consistent with the estimates, where a consistency test is defined below.
Similarly to the edges parallel to the moving direction, there can be multiple edges of an estimated length and direction.
Thus, in (iv), we estimate the number of edges the length and direction of which are $\widehat{\lambda},\widehat{\xi}$.

\subsubsection{Temporary estimation}
For temporary estimation, we use two pairs of measured sensing results $(\senseResult l j i, \senseResult s j i), (\senseResult l {j'} {i'}, \senseResult s {j'} {i'})\not\in \Psi(s_d=0)$.
Apply Section \ref{two-sol} to these two pairs, use $\widehat{v}$ as $v$, and obtain the temporary estimates $\tilde{\lambda},\tilde{\xi}$.
If these two pairs of measured sensing results are those of the same edge, the temporary estimates should be good estimates.
Otherwise, they are meaningless.
Therefore, we should choose sensing result pairs that are likely to be sensing results of the same edge.

To efficiently find such pairs, we should classify $\{(\senseResult l j i, \senseResult s j i)\}_{i,j}\not \in \Psi(s_d=0)$ into several sets.
If $(\senseResult l j i, \senseResult s j i)$ and $(\senseResult l {j'} {i'}, \senseResult s {j'} {i'})$ are the sensing results regarding the same edge, they are likely to belong to the same set defined below.
Let $\Psi(0<s_d\ll 1)$, $\Psi(-1\ll s_d<0)$, $\Psi(1\ll s_d)$, $\Psi(s_d\ll -1)$, $\Psi(s_d\approx 1)$, and $\Psi(s_d\approx -1)$ be examples of such sets.
$(\senseResult l j i, \senseResult s j i)$ belonging to $\Psi(\gamma)$ means that $\senseResult s j i$ satisfies $\gamma$ where $\gamma\in \{0<s_d\ll 1,-1\ll s_d<0,1\ll s_d,s_d\ll -1,s_d\approx 1,s_d\approx -1\}$.
For the following reasons, these sets are good candidates for sets classifying $\{(\senseResult l j i, \senseResult s j i)\}_{i,j}$.

According to Eq. (\ref{abs_sin_xi}), $|s_d|\ll 1$ means $|\sin\xi|\ll 1$.
That is, $\xi\approx 0,\pi$.
Because of Eq. (\ref{sdsinxi}), $\Psi(0<s_d\ll 1)$ corresponds to $\{\xi\approx 2\pi$ or $\pi\}$ and $\{\xi\in(\pi,2\pi)\}$.
Similarly, $\Psi(-1\ll s_d<0)$ corresponds to $\{\xi\approx 0$ or $\pi\}$ and $\{\xi\in(0,\pi)\}$.
Thus, if we choose $(\senseResult l j i, \senseResult s j i)$ and $(\senseResult l {j'} {i'}, \senseResult s {j'} {i'})$ in $\Psi(0<s_d\ll 1)$ ($\Psi(-1\ll s_d<0)$), it is likely that those two pairs are sensing results of the same edge.

Due to Eq. (\ref{s_d}), $|s_d|\gg 1$ means $\theta\approx\xi,\xi+\pi$.
Then, because of Eq. (\ref{ldsd}), $l_d|s_d|\approx \lambda/v$.
Therefore, we use $\Psi(1\ll s_d)$ ($\Psi(s_d\ll -1)$) and, if needed, divide $\Psi(1\ll s_d)$ ($\Psi(s_d\ll -1)$) into subsets of elements that have similar $l_d|s_d|$.

The remaining pairs $\{(\senseResult l j i, \senseResult s j i)\}_{i,j}$ that do not belong to any of sets mentioned above belong to sets $\Psi(s_d\approx 1)$
$\Psi(s_d\approx -1)$.

\subsubsection{Consistency test}
If $l_d$ and $s_d$ are sensing results for the edge the length and direction of which are exactly $\tilde{\lambda}$ and $\tilde{\xi}$, Eq. (\ref{xi-sol}) with $\lambda=\tilde{\lambda},v=\widehat{v}$ yields $\tilde{\xi}$.
In accordance with this fact, the consistency test of a pair $l_d,s_d$ is defined as the following procedure for a given temporary estimates $\tilde{\lambda},\tilde{\xi}$:

\begin{quote}
Compute Eq. (\ref{xi-sol}) with $\lambda=\tilde{\lambda}$ and $v=\widehat{v}$ to obtain $\xi(l_d,s_d)$ and compare this $\xi(l_d,s_d)$ with $\tilde{\xi}$.
(Alternatively, compute Eq. (\ref{xi-by-lam}) with $\lambda=\tilde{\lambda}$ and $v=\widehat{v}$ to obtain $\mu$, and compare this $\mu$ with $\cos\tilde{\xi}$.)
If they (approximately) agree, this pair $l_d,s_d$ passes the test.
That is, this pair $(l_d,s_d)$ is consistent with the temporary estimates $\tilde{\lambda},\tilde{\xi}$.
Otherwise, it fails.
\end{quote}

\subsubsection{Adoption as estimates}
The first pair we adopt as an estimate pair $(\widehat{\lambda},\widehat{\xi})$ is the temporary estimate pair $(\tilde{\lambda},\tilde{\xi})$ that has the largest number of consistent sensing results in $\Psi_1(s_d\neq 0)$ where $\Psi_1(s_d\neq 0)$ is the set of sensing results $\{(\senseResult l j i, \senseResult s j i)\}_{i,j}$ not included in $\Psi(s_d=0)$.
In general, the $k$-th estimate pair we adopt is the estimate pair that has the largest number of consistent sensing results in $\Psi_k(s_d\neq 0)$.
Let $ \Psi_{k,0}(s_d\neq 0)$ be the sensing results consistent with the $k$-th estimate pairs.
By removing $ \Psi_{k,0}(s_d\neq 0)$ from $\Psi_k(s_d\neq 0)$, define $\Psi_{k+1}(s_d\neq 0)$.

\subsubsection{Estimating number of edges}
There can be one or more edges the length and direction of which are $\widehat{\lambda},\widehat{\xi}$.
Similar to Section \ref{part0direction}, the number of edges the length and direction of which are $\widehat{\lambda},\widehat{\xi}$ can be estimated through $E[n_d(\lambda,\xi)]$ given by Eq. (\ref{detect_num}) with $\lambda=\widehat{\lambda},\xi=\widehat{\xi}$. 
We use $\widehat{n_e}(\widehat{\lambda},\widehat{\xi})$ as the estimate of the number of edges the length and direction of which are $\widehat{\lambda},\widehat{\xi}$.
The number of estimated edges through this method can be zero when $\widehat{n_e}(\widehat{\lambda},\widehat{\xi})$ is small.

\subsection{Fourth part: Estimating order of edges}
The second and third parts provide us pairs of edge length and direction.
However, to identify the shape of the target, we need to identify consecutive edges, that is, the order of edges that connect.

To derive a method identifying the order of edges, pay attention to the behavior of $r(t)$.
$r(t)$ is continuous and becomes two consecutive line segment parts $p_d(L_{j}|\theta),p_d(L_{j+1}|\theta)$ when a sensor the direction of which is $\theta$ detects consecutive edges $L_{j},L_{j+1}$.
We use data detecting the whole of consecutive edges $L_{j},L_{j+1}$, but we do not know $j$.

Assume that $(\senseResult l m i, \senseResult s m i)$ and $(\senseResult l {m+1} i, \senseResult s {m+1} i)$ are the $i$-th sensor's consecutive sensing results and belong to $\Psi_{a,a'}$ and $\Psi_{b,b'}$ where $\Psi_{a,a'},\Psi_{b,b'}\in \{\Psi_{k,0}(s_d\neq 0)\}_k\cup \{\Psi_{0,k}(s_d=0)\}_k$.
If a sensor consecutively detects multiple edges without jumps of $r(t)$, these edges are consecutive.
If these sensing results are consistent with $(\widehat{\lambda}(\Psi_{a,a'}),\widehat{\xi}(\Psi_{a,a'}))$ and $(\widehat{\lambda}(\Psi_{b,b'}),\widehat{\xi}(\Psi_{b,b'}))$, respectively, it is likely that an edge the length and direction of which are $\widehat{\lambda}(\Psi_{a,a'}),\widehat{\xi}(\Psi_{a,a'})$ connects to an edge the length and direction of which $\widehat{\lambda}(\Psi_{b,b'}),\widehat{\xi}(\Psi_{b,b'})$.
Let $n_c(a,a';b,b')$ be the number of consecutive sensing results belonging to $\Psi_{a,a'}$ and $\Psi_{b,b'}$.
We judge that an edge of length $\widehat{\lambda}(\Psi_{a,a'})$ and direction $\widehat{\xi}(\Psi_{a,a'})$ connects to an edge of length $\widehat{\lambda}(\Psi_{b,b'})$ and direction $\widehat{\xi}(\Psi_{b,b'})$, if $n_c(a,a';b,b')$ is large.

When the $i$-th sensor's sensing results $(\senseResult l m i, \senseResult s m i)$ and $(\senseResult l {m+1} i, \senseResult s {m+1} i)$ are consecutive and $\senseResult s m i>\senseResult s {m+1} i$, this vertex formed by two consecutive edges for which sensing results are $(\senseResult l m i, \senseResult s m i)$ and $(\senseResult l {m+1} i, \senseResult s {m+1} i)$ is concave.
Otherwise, it is convex.
Because the temporary estimates in the second and third parts based on Section \ref{two-sol} are not unique, this information regarding convexity/concavity is useful to reduce the number of combinations of estimates.

\begin{figure}[tb] 
\begin{center} 
\includegraphics[width=9cm,clip]{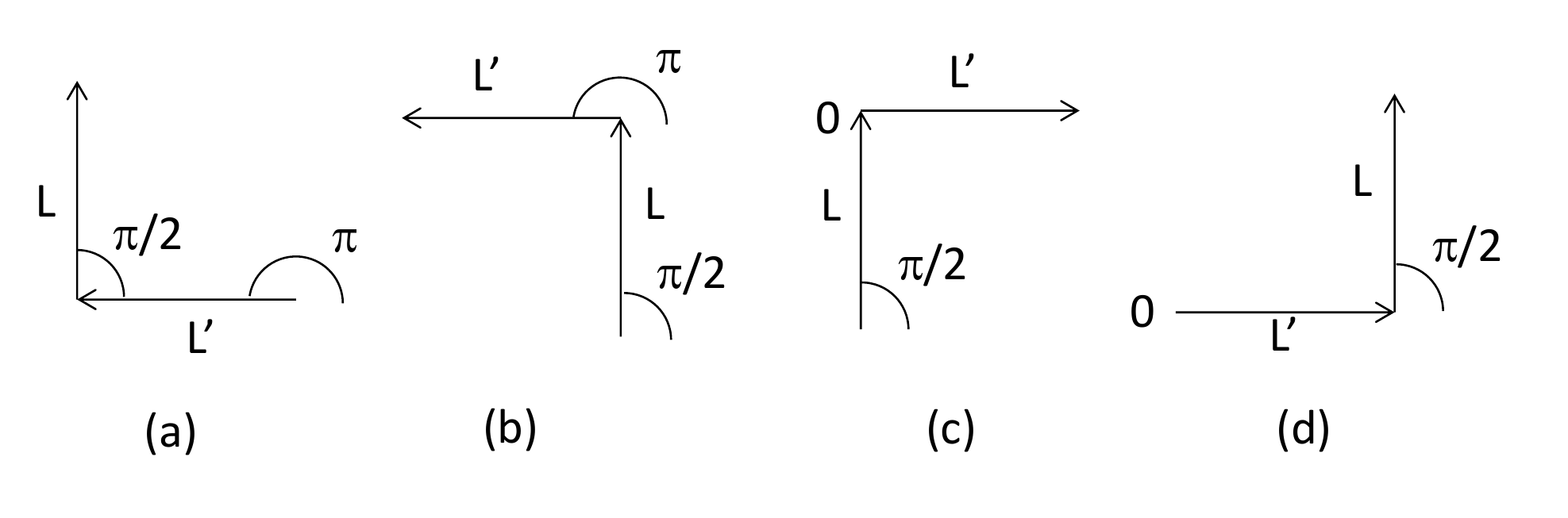} 
\caption{Example of patterns connecting two edges} 
\label{example_connect} 
\end{center} 
\end{figure}

In addition, the order of detection by a single sensor provides the locations of edges.
When $T$ moves right, an edge to the right is detected earlier than that to the left by a single sensor.
This information is particularly useful to reduce the number of patterns of connected edges.
For example, when the direction of an edge $L$ is estimated as $\pi/2$ and its consecutive edge $L'$ is estimated as 0 or $\pi$, their connectivity patterns are illustrated in Fig. \ref{example_connect}.
When $T$ moves right and a sensor detects $L$ first and detects $L'$ later, we can conclude that $L$ and $L'$ connect as (b) or (d), not (a) or (c).
This conclusion is independent of the sensor direction.

\subsection{Fifth part: Compensating edges forming concave vertex}
This part may provide additional estimates of edges forming a concave vertex of $T$.
(By finding jumps in $r(t)$, we can judge the existence of a concave vertex.) 
As described below, the estimated number of edges with $E[n_d(\lambda,\xi)]$ given by Eq. (\ref{detect_num}) in the second and third parts may underestimate the number of edges for non-convex $T$.
As an extreme case, the estimated number becomes zero.
This part compensates for this error.

When $T$ has a concave part, $E[n_d(\lambda,\xi)]$ may not be given by Eq. (\ref{detect_num}).
The reason is that a sensor that should detect this edge may not because a part of $T$ is between this edge and this sensor.
That is, a part of $T$ blocks this sensor from detecting this edge.

\begin{figure}[tb] 
\begin{center} 
\includegraphics[width=9cm,clip]{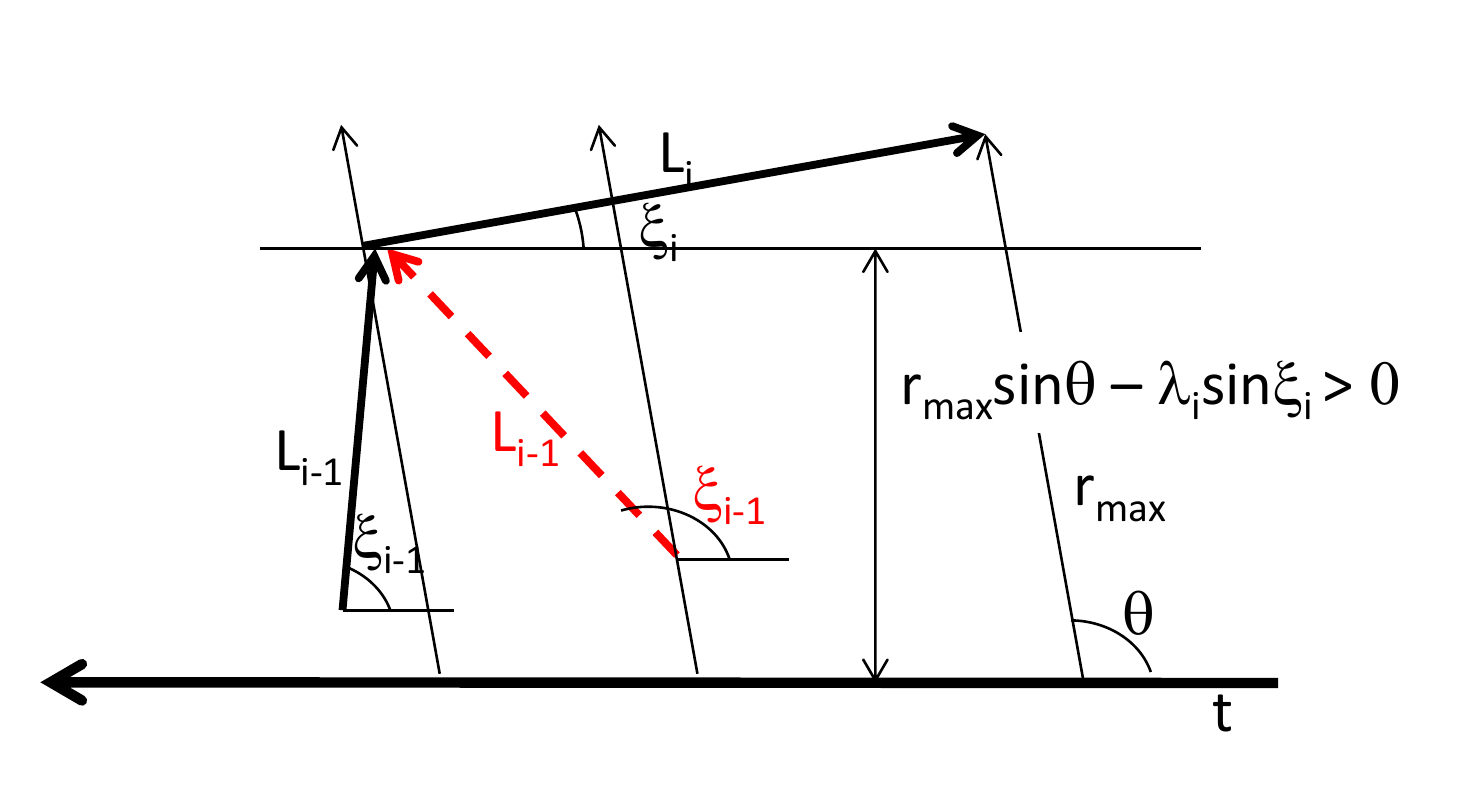} 
\caption{Blocking detection of $L_i$ due to convexity} 
\label{concave} 
\end{center} 
\end{figure}

We take account of this block and modify Eq. (\ref{detect_num}) for non-convex $T$.
We consider an event in which two consecutive edges form a concave vertex of $T$ and one may block the detection of the other.
We neglect other blocking events caused by other edges.
As shown in Fig. \ref{concave}, the detection of $L_i$ by the sensor the direction of which is $\theta$ is not blocked only if $\xi_{i-1}<\theta$.
In addition, similarly to Section \ref{num_result}, to detect the whole $L_i$ (if no blocking), $\theta\in\mod{\xi_i}{\xi_{i}+\pi}$ and  $\theta\in[\theta_{min1},\pi-\theta_{min1})\cup[\pi+\theta_{min1},2\pi-\theta_{min1})$.
Hence, $E[n_d(\lambda_i,\xi_{i})]$ for a concave vertex ($\xi_{i-1}\in\mod{\xi_{i}} {\xi_{i}+\pi}$) with the condition $\lambda|\sin\xi_{i}|\leq r_{max}$ is given as follows.
\bqn
&&E[n_d(\lambda_i,\xi_{i})]\cr
&=&vm_tn_s\frac{\int_{\theta\in\Theta(\theta_{min1})}(r_{max}|\sin\theta|-\lambda|\sin\xi_i|) d\theta}{2\pi\sizex{\Omega}}\cr
&=&vm_tn_sf(\theta_{min1},\lambda|\sin\xi_i|)/(2\pi\sizex{\Omega}),\label{concave_n_d}
\eqn
where $\Theta(\theta)\defeq ([\theta,\pi-\theta)\cup[\pi+\theta,2\pi-\theta))\cap[\xi_{i-1},\xi_i+\pi)$ and 
\bq
f(\theta,x)\defeq\int_{z\in\Theta(\theta)}(r_{max}|\sin(z)|-x) dz.
\eq
$f(\theta,x)$ is given in Appendix.

Instead of Eq. (\ref{detect_num}), use Eq. (\ref{concave_n_d}) for concave vertexes and reevaluate $\widehat{n_e}(\widehat{\lambda},\widehat{\xi})$ in the second and third parts.
Reevaluation may result in an increase in the number of edges.

\section{Numerical examples}
\subsection{Default conditions of examples}
In this section, the following conditions are used as the default conditions unless explicitly indicated otherwise.

$\Omega$ is a rectangular area of $5000\times 300$ the longer edge of which is along the $x$-axis.
$T$ is moving on the centerline parallel to the $x$-axis of $\Omega$.
Although our model was continuous in time, sensors report at a single time unit interval in the simulation.
$r_{max}=100$, $n_s=2000$, $v=1$.
If $|r(t+1)-r(t)|<0.1$, we judge that $s_d=0$.
As a consistency test, $\mu$ derived by Eq. (\ref{xi-by-lam}) was used and it was tested that $|\cos\tilde{\xi}|$ was between $\mu(\lambda=0.85\tilde{\lambda})$ and $\mu(\lambda=1.15\tilde{\lambda})$.

\subsection{Basic examples}\label{basic_ex}
To understand the behavior of the estimates derived by the proposed method, we use a simple $T$.
This is a triangle of $\{(\lambda_i,\xi_i)\}_{i=1,2,3}=(50\sqrt{3},0),(100,5\pi/6),(50,3\pi/2)$.
Because a triangle has a unique order of edges connecting except for its mirror image, we can evaluate the shape estimation accuracy by the accuracy of the estimated length and direction of each edge without taking account of the order of edges connecting.
Therefore, two metrics are evaluated by using ten simulation runs for each case:
(i) mean square error (MSE) $\defeq\sum_{j=1}^{10}\sum_{i=1}^3 \epsilon_{i,j}^2/10$, and (ii) relative square root of MSE for $L_i$ (RSR-MSE$_i$) $\defeq\sqrt{\sum_{j=1}^{10}\epsilon_{i,j}^2/10}/\lambda_i$.
Here, $\epsilon_{i,j}^2$ is an square error of the estimated location of the head of $L_i$ at the $j$-th simulation run when its tail is placed at the origin.
That is, $\epsilon_{i,j}^2\defeq (\lambda_i\cos\xi_i-\widehat{\lambda_i}\cos\widehat{\xi_i})^2+(\lambda_i\sin\xi_i-\widehat{\lambda_i}\sin\widehat{\xi_i})^2$ derived at the $j$-th simulation.
(For two estimates of $\xi_i$, the estimate minimizing the square error is adopted as the formal estimate.)
Although the number of edges estimated may not be three for some ill conditions, these metrics are calculated for the first three estimates obtained.

\begin{figure}[tb] 
\begin{center} 
\includegraphics[width=9cm,clip]{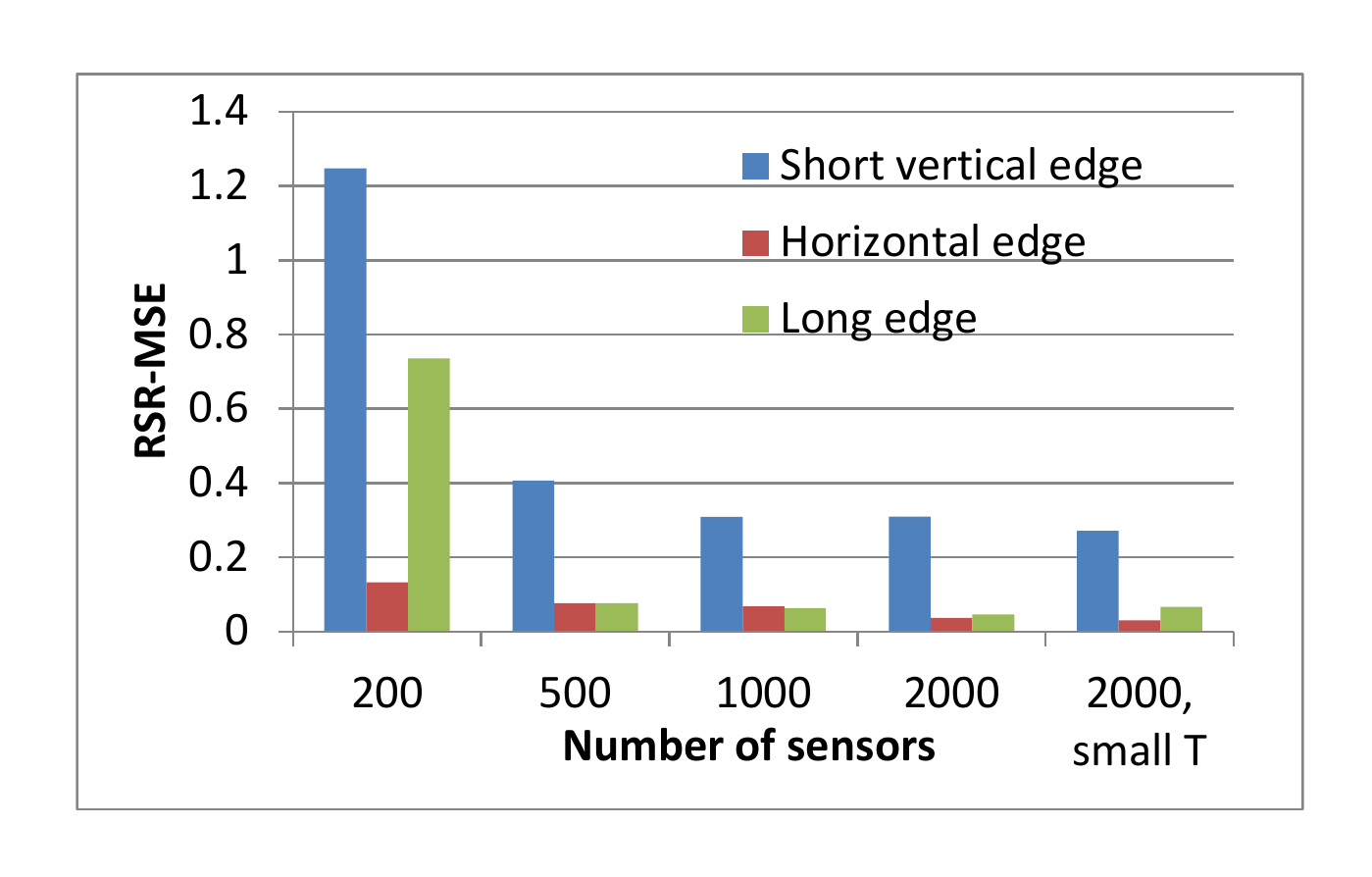} 
\caption{Relationship between number of sensors and estimation accuracy} 
\label{RSR-MSE} 
\end{center} 
\end{figure}

First, the relationship between the number of sensors and the estimation accuracy was investigated.
Fig. \ref{RSR-MSE} plots RSR-MSE for each edge.
Additionally, a small $T$ that has edges half as long as those of the original $T$ was also used to evaluate the estimation accuracy.
There were three important findings. 
(1) A vertical edge (short edge) was very difficult to estimate whereas a horizontal edge was easy.  The former caused errors of about 30\% and the latter caused errors of a few percent for $n_s\geq 1000$.
(2) The estimation accuracy is fairly insensitive to the size of $T$.  
Therefore, the vertical edge (short edge) seems to be difficult to estimate mainly because it is vertical not because it is short.
(3) The number of sensors $n_s$ should be larger than 500, but the estimation accuracy is fairly insensitive to $n_s$ if $n_s>500$.

\begin{figure}[tb] 
\begin{center} 
\includegraphics[width=9cm,clip]{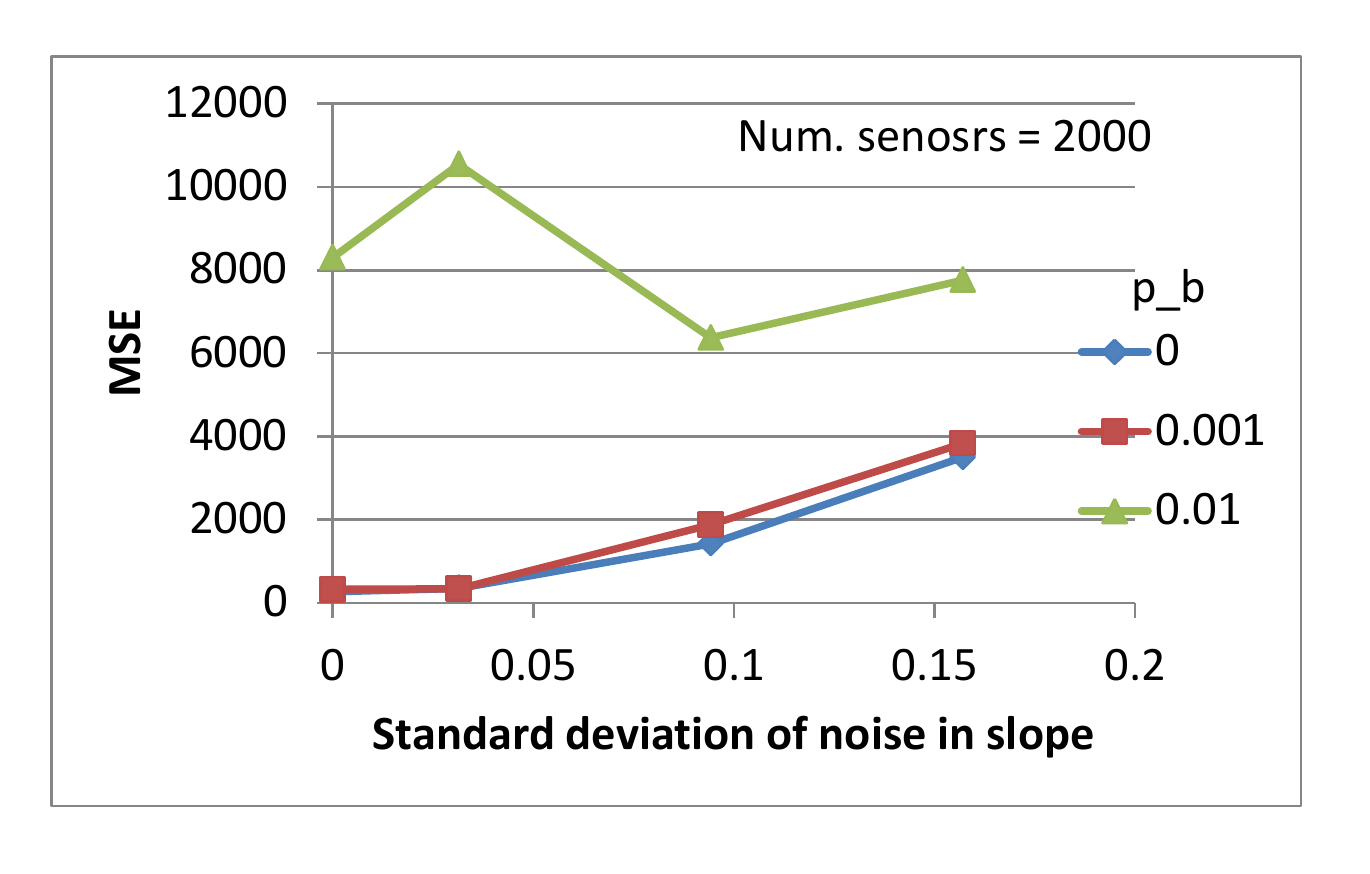} 
\caption{Impact of noise on estimation accuracy} 
\label{noise_edge} 
\end{center} 
\end{figure}

Second, the impact of noise on the estimation accuracy was investigated.
For the set of observed periods $p_d$ without noise, noises were imposed.
With probability $p_b$, $r(t)$ was lost at each $t$.
As a result, $l_d$ for this $p_d$ was broken at this $t$.
In addition, a zero-mean Gaussian noise was imposed on each $s_d$.
Figure \ref{noise_edge} plots the MSE.
For $l_dp_b\ll 1$ such as $p_b=0,0.001$, the MSE super-linearly increased as the standard deviation of the noise on $s_d$ increased.
When $l_dp_b\approx 1$ such as $p_b=0.01$, the estimation became very poor for any standard deviation of the noise on $s_d$.

\begin{figure}[tb] 
\begin{center} 
\includegraphics[width=9cm,clip]{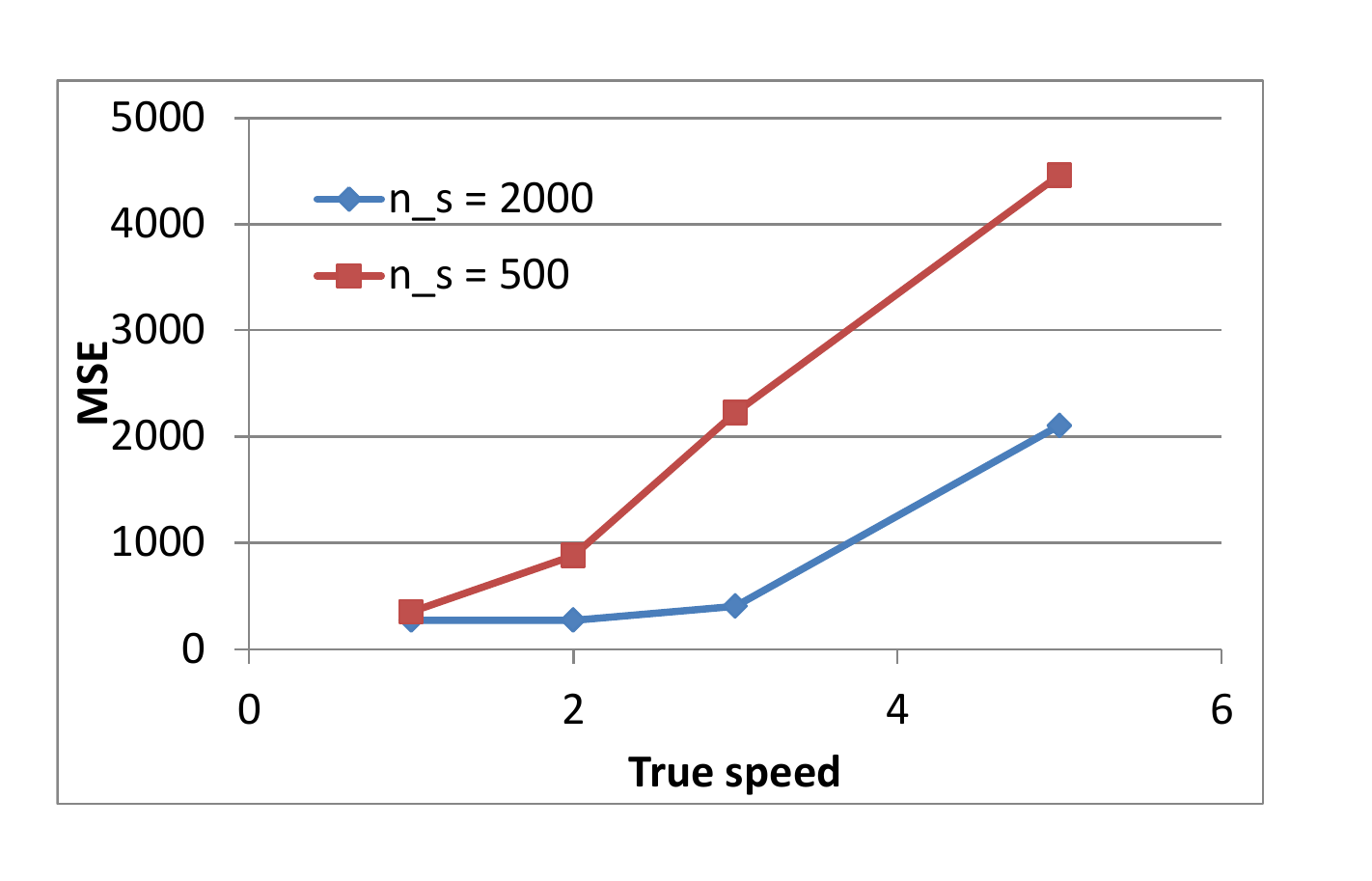} 
\caption{Relationship between target object moving speed and estimation accuracy} 
\label{speed_edge} 
\end{center} 
\end{figure}

Third, the estimation accuracy for various speeds $v$ is investigated.
As demonstrated in Figure \ref{speed_edge}, the shape of $T$ is difficult to estimate when $T$ moves fast.
This seems to be because we use sensor reports at a single time unit interval.
Therefore, when a sensor finishes detecting one edge and and starts detecting another, we may miss the exact epoch of this change in the edges detected.
This can introduce a sensing error.
For a small $n_s$, MSE is sensitive to $v$.

\subsection{Realistic examples}
\begin{figure}[tb] 
\begin{center} 
\includegraphics[width=9cm,clip]{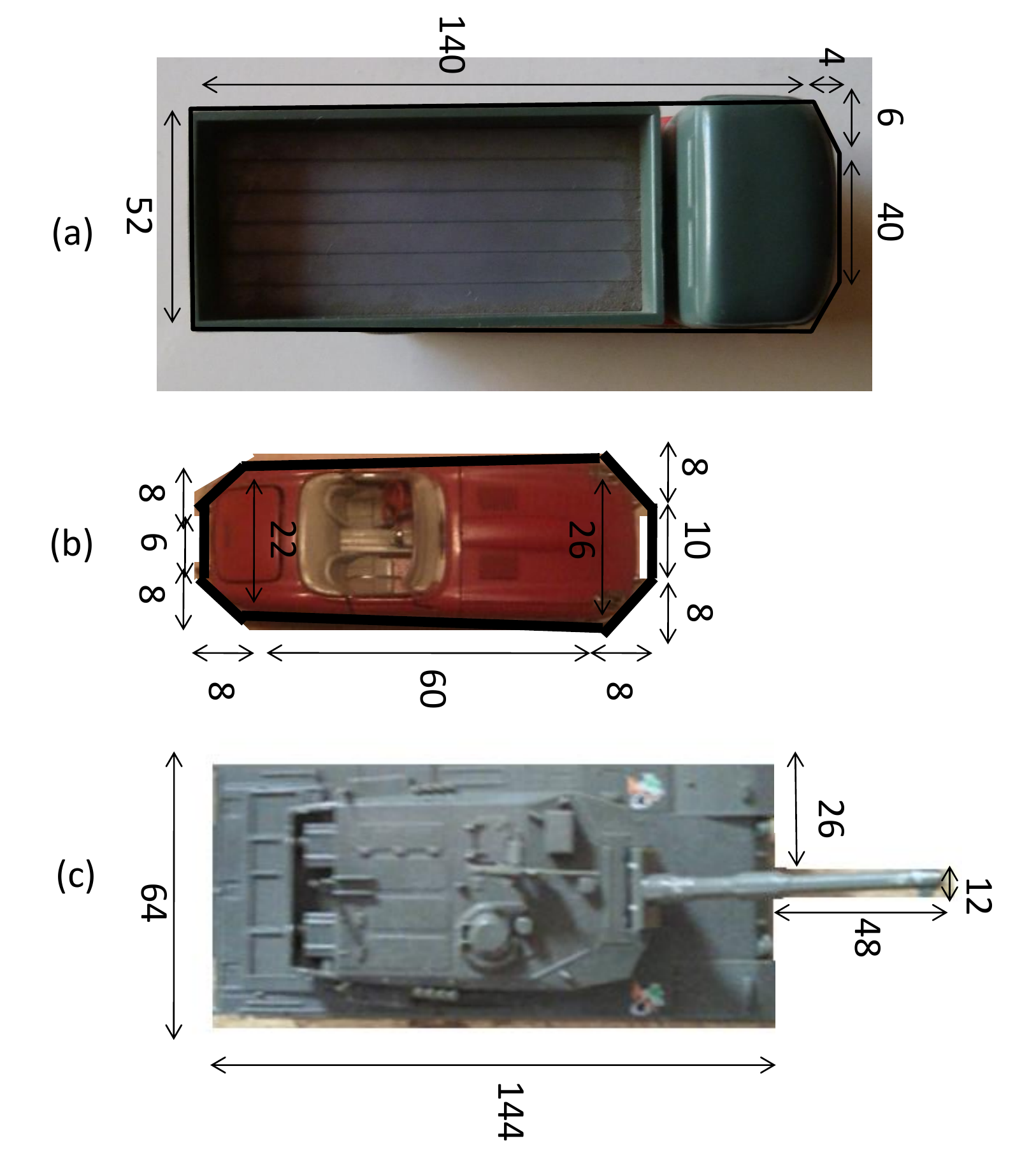} 
\caption{Examples of $T$} 
\label{target} 
\end{center} 
\end{figure}

We applied the proposed estimation method to three toy-vehicles ((a) truck, (b) sports car, and (c) tank) shown in Fig. \ref{target}.
The truck and sports car are convex, and the truck and tank have edges along the moving direction.
For these examples, the number of edges of length $\widehat{\lambda}(\Psi_{a,a'})$ and direction $\widehat{\xi}(\Psi_{a,a'})$ is estimated as $\lfloor\widehat{n_e}(\widehat{\lambda}(\Psi_{a,a'}),\widehat{\xi}(\Psi_{a,a'}))+0.5\rfloor$.
Even if this number is zero, we set this number as one if there exists $(b,b')$ such that $n_c(a,a';b,b')\geq 30$.

In the following examples, in addition to the mirror image of an estimated shape, ambiguity caused by two estimates of $\xi$ exists and the estimated shape of $T$ cannot be uniquely determined.
However, the ambiguity of its shape is fairly small.
In addition, note that both ends of $\partial T$ may not meet.

\subsubsection{Shape estimation of truck}
In accordance with the second and third parts of our proposed method, we obtained the estimated edge length and direction shown in Table \ref{est_list_truck}.
We also obtained consecutive edges in the fourth part.
The obtained results are shown in Table \ref{est_list_truck_con} where two consecutive edges are listed in a line.
Among two consecutive edges, an edge closer to the head of truck is shown in the ^^ ^^ Head" column and an edge closer to the tail is shown in the  ^^ ^^ Tail" column.
Edge IDs (i) - (iv) used were those in Table \ref{est_list_truck}.

\begin{figure}[tb] 
\begin{center} 
\includegraphics[width=9cm,clip]{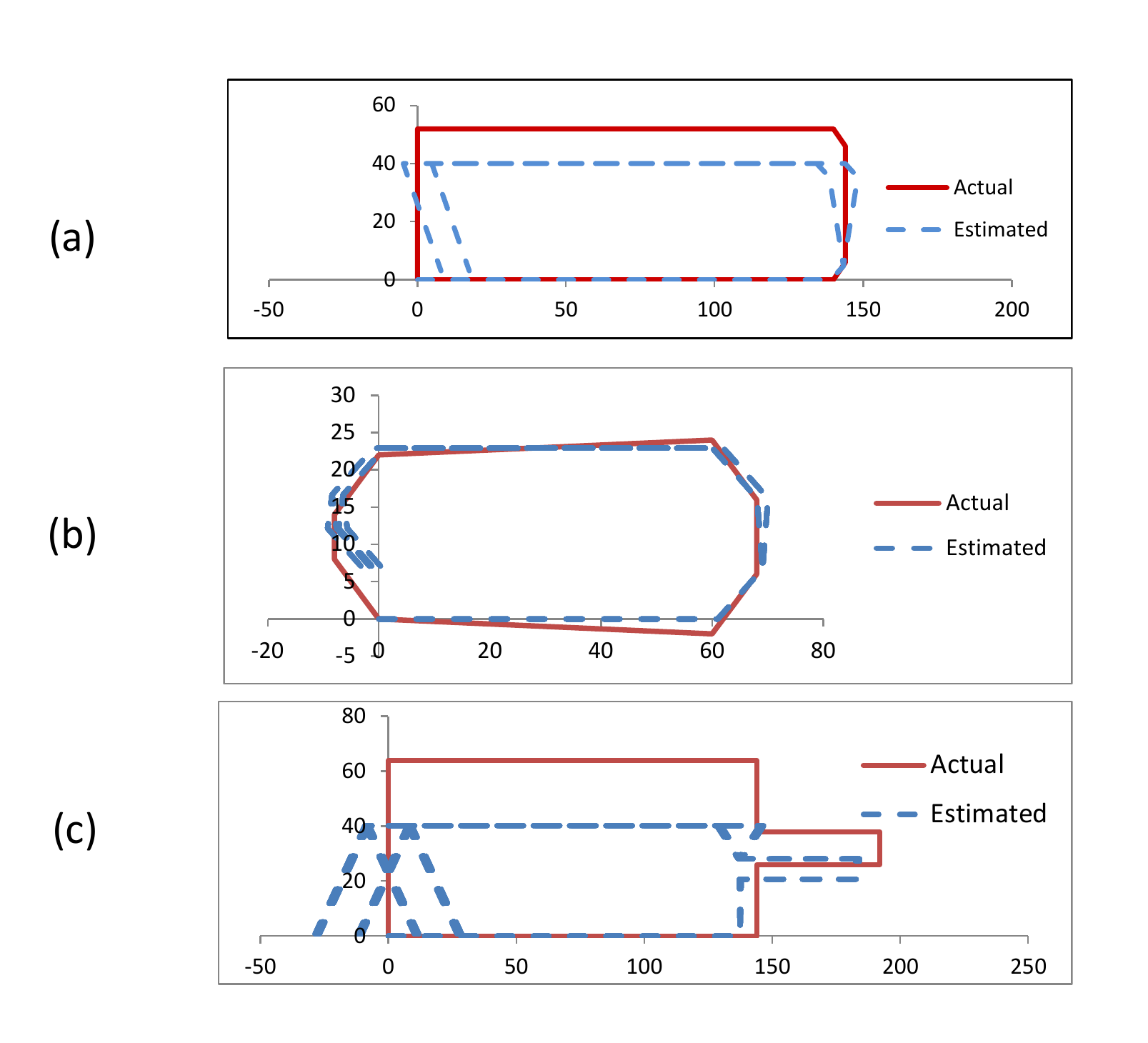} 
\caption{Estimated shape of target objects} 
\label{target_estimated} 
\end{center} 
\end{figure}

Results in these tables enabled us to plot the shape of $T$.
For example, a near-vertical edge (iii) connects to short two edges (ii), each of (ii) connects to a horizontal edge (i), and two edges (i) connect to edge (iv).
The fifth step in the proposed method was not applied.
The estimated shape of $T$ is plotted in Fig. \ref{target_estimated}-(a).
(Here, the shapes where $\partial T$ is almost complete are plotted.
There are combinations of $\xi_i$ estimated as shown in Table \ref{est_list_truck} that do not make $\partial T$ at all.
For example, if both estimated edges (i) take the direction 0, we cannot make $\partial T$ at all.)

The estimated shape was slightly more slender than the actual shape, and the vertical edges were not estimated as vertical.
That is, the error of the shape estimation mainly comes from the estimation error of vertical edges.
This is consistent with the results for the basic examples.

\begin{table}
\caption{Estimated edge length and direction of truck}
\begin{center}\label{est_list_truck}
\begin{tabular}{llll}
\hline
&Num. edges&$\widehat{\lambda}$&$\widehat{\xi}$\\
(i)&2&139.1&0,$\pi$\\
(ii)&2&6.348&0.8355,2.306\\
(iii)&1&31.03&1.726,1.416\\
(iv)&1&39.10&-1.900,-1.241\\
\hline
\end{tabular}
\end{center}

\caption{Estimated consecutive edges of truck}
\begin{center}\label{est_list_truck_con}
\begin{tabular}{lll}
\hline
Head&Tail&Num. samples\\
(iii)&(ii)&104\\
(ii)&(i)&252\\
(i)&(iv)&70\\
\hline
\end{tabular}
\end{center}
\end{table}

\subsubsection{Shape estimation of sports car}
Similar to the example of the truck, the proposed estimation was applied to the sports car.
Tables \ref{est_list_car} and \ref{est_list_car_con} were obtained.
Again, the fifth step in the proposed method was not applied.

The estimated shape plotted in Fig. \ref{target_estimated}-(b) looks similar to the actual shape.
The estimation errors mainly occurred for the following two reasons.
One was the error of one vertical edge: its estimated length was too short.
The other was that the two long nearly horizontal edges were estimated as horizontal edges.
Because they are long, small errors in direction resulted in large errors in the estimated shape.

\begin{table}
\caption{Estimated edge length and direction of sports car}
\begin{center}\label{est_list_car}
\begin{tabular}{llll}
\hline
&Num. edges&$\widehat{\lambda}$&$\widehat{\xi}$\\
(i)&2&61.02&0,$\pi$\\
(ii)&2&10.51&0.6965,2.445\\
(iii)&2&9.731&-2.439,-0.7022\\
(iv)&1&9.507&1.467,1.674\\
(v)&1&4.569&-1.399,-1.742\\
\hline
\end{tabular}
\end{center}

\caption{Estimated consecutive edges of sports car}
\begin{center}\label{est_list_car_con}
\begin{tabular}{lll}
\hline
Head&Tail&Num. samples\\
(iv)&(ii)&83\\
(ii)&(i)&305\\
(i)&(iii)&280\\
(iii)&(iii)&51\\
(iii)&(v)&30\\
\hline
\end{tabular}
\end{center}
\end{table}

\subsubsection{Shape estimation of tank}
In accordance with our proposed method, we obtained the estimated edge length and direction shown in Table \ref{est_list_tank} where the number of edges in parentheses means that after applying the fifth part of our proposed method.
We also obtained consecutive edges in Table \ref{est_list_tank_con}.
The shape was plotted in Fig. \ref{target_estimated}-(c).
Connections between edge (iii) near the head and edge (i) near the tail were observed but not used in the estimated shape in $T$.
This seems to be because the proposed method did not accurately distinguish short vertical edges.
That is, the connection between edges (iii) and (i) should have been that between edges (iv) and (i).

The estimated shape was too slender.
The reason for the estimation error in the shape of $T$ seemed similar to that for the first example (truck).
That is, the estimation error of vertical edges was the main reason.

All through these three examples, edge lengths tend to be underestimated. A possible reason is that $r(t)$ is not exactly
continuous but sampled at every one unit time interval. This sampling may cut the end of a detected edge and may result in a short 
edge length.
Especially, this estimation error seems to be more serious to vertical edges than to other ones.
This is probably because as for a vertical edge, the detection time length $l_d$ is typically short
and the change of the distance $|r(t_e)-r(t_s)|$ is large as explained below.

Measurement errors include the detection time error $\Delta l_d$ and the slope error $\Delta s_d$. While the former appears
systematically with negative value as explained above and should be considered, the latter is not serious because the slope $s_d$
is not so sensitive to the time sampling.
Thus, the magnitude of edge length's error mainly depends on the derivative of $\lambda$ with respective to $l_d$, not $s_d$ as follows.

\bqn
\frac{\partial\lambda}{\partial l_d}=\frac{1}{2\lambda(1-x)}\left\{2\left[\frac{(r(t_e)-r(t_s))^2}{l_d}-l_d\right] \right. \\
\left. -\left[\frac{(r(t'_e)-r(t'_s))^2}{l'_d}-l'_d\right]+\frac{\lambda^2}{l'_d}\right\},
\label{eq}
\eqn
where $x=l_d/l'_d$.

With $\lambda$ held constant, Eq.~(\ref{eq}) appears to be large when $l_d$ is small and $|r(t_e)-r(t_s)|$ is large.
Eq.~(\ref{eq}) depends on $\xi$ implicitly through $l_d$ and $|r(t_e)-r(t_s)|$.
In fact, it can be shown that the expectation of $l_d$ under the measure
$(r_max|\mathrm{sin}\theta|-\lambda|\mathrm{sin}\xi|)d\theta/
\int_{\Theta_0}(r_max|\mathrm{sin}\theta|-\lambda|\mathrm{sin}\xi|)d\theta$ monotonously decreases with respect to $\xi$
over the interval $[0,\pi/2]$, and that of $|r(t_e)-r(t_s)|$ monotonously increases over the same interval.
This fact supports the hypothesis that the estimation error of vertical edge ($\xi=\pi/2$) length is likely to be large because of
small $l_d$ and large $|r(t_e)-r(t_s)|$.

\begin{table}
\caption{Estimated edge length and direction of tank}
\begin{center}\label{est_list_tank}
\begin{tabular}{llll}
\hline
&Num. edges&$\widehat{\lambda}$&$\widehat{\xi}$\\
(i)&1(2)&45.84&0,$\pi$\\
(ii)&2&137.5&0,$\pi$\\
(iii)&1&14.70&0.9678,2.174\\
(iv)&1&7.420&1.627,1.514\\
(v)&1&44.00&-1.113,-2.028\\
(vi)&1&20.69&1.591,1.550\\
\hline
\end{tabular}
\end{center}

\caption{Estimated consecutive edges of tank}
\begin{center}\label{est_list_tank_con}
\begin{tabular}{lll}
\hline
Head&Tail&Num. samples\\
(iv)&(i)&78\\
(i)&(iii)&99\\
(iii)&(ii)&113\\
(iii)&(i)&80\\
(ii)&(v)&55\\
(i)&(vi)&30\\
(vi)&(ii)&31\\
\hline
\end{tabular}
\end{center}
\end{table}

\section{Conclusion}
This paper proposed a method for estimating the shape of a target object moving at an unknown speed and unknown location by using location-unknown sensors.
This proposed method demonstrated that simple sensors without location information can estimate a target-object shape even though there are many unknown factors.
The estimate may not be accurate enough, but the proposed method presents a new direction for shape estimation.
Simultaneously, this method is important as a crowdsensing and participatory sensing that maintains location privacy.

The following remain as for further study.
(1) The proposed method used sensing data that detected whole edges.
This means that some sensing data were not used.
Therefore, sensing data that do not correspond to whole edges need to be efficiently used.
(2) The proposed method assumed the polygon target object and moving on a straight line.
It thus needs to be extended to a non-polygon target object and a non-straight line movement.

In addition to a theoretical study, an experiment using the proposed method also remains as further study.

\appendix
For $\theta\in [0,\pi/2]$, define $Zone_1(\theta)\defeq [0,\theta)\cup[2\pi-\theta,2\pi)$, $Zone_2(\theta)\defeq [\theta,\pi-\theta)$, $Zone_3(\theta)\defeq [\pi-\theta,\pi+\theta)$, $Zone_4(\theta)\defeq [\pi+\theta,2\pi-\theta)$.
For $0\leq\xi_{i-1},\xi_i<2\pi$, $f(\theta,x)$ is given as follows for $\xi_{i-1}\in[\xi_i,\xi_i+\pi]$.

When $\xi_{i-1},\xi_i$ are in $Zone_1(\theta)$, $\Theta =[\theta, \pi-\theta)$.  Thus, $f(\theta,x)=2r\cos\theta-x(\pi-2\theta)$.

When $\xi_i$ is in $Zone_1(\theta)$ and $\xi_{i-1}$ is in $Zone_2(\theta)$, $\Theta =[\xi_{i-1}, \pi-\theta)$.
Thus, $f(\theta,x)=r(\cos\xi_{i-1}+\cos\theta)-x(\pi-\theta-\xi_{i-1})$.

When $\xi_i$ is in $Zone_1(\theta)$ and $\xi_{i-1}$ is in $Zone_3(\theta)$, $\Theta =\emptyset$.
Thus, $f(\theta,x)=0$.

When $\xi_{i-1},\xi_i$ are in $Zone_2(\theta)$, $\Theta =[\xi_{i-1},\pi-\theta)\cup[\pi+\theta,\pi+\xi_i)$.
Thus, $f(\theta,x)=r(2\cos\theta+\cos\xi_{i-1}-\cos\xi_i)-x(\pi-2\theta-\xi_{i-1}+\xi_i)$.

When $\xi_i$ is in $Zone_2(\theta)$ and $\xi_{i-1}$ is in $Zone_3(\theta)$, $\Theta =[\pi+\theta,\pi+\xi_i)$.
Thus, $f(\theta,x)=r(-\cos\xi_i+\cos\theta)-x(\xi_i-\theta)$.

When $\xi_i$ is in $Zone_2(\theta)$ and $\xi_{i-1}$ is in $Zone_4(\theta)$, $\Theta =[\xi_{i-1},\xi_i+\pi)$.
Thus, $f(\theta,x)=-r(\cos\xi_{i-1}+\cos\xi_i)-x(-\xi_{i-1}+\xi_i+\pi)$.

When $\xi_{i-1},\xi_i$ are in $Zone_3(\theta)$, $\Theta =[\theta+\pi, 2\pi-\theta)$.
Thus, $f(\theta,x)=2r\cos\theta-x(\pi-2\theta)$.

When $\xi_i$ is in $Zone_3(\theta)$ and $\xi_{i-1}$ is in $Zone_4(\theta)$, $\Theta =[\xi_{i-1},2\pi-\theta)$.
Thus, $f(\theta,x)=r(\cos\theta-\cos\xi_{i-1})-x(2\pi-\theta-\xi_{i-1})$.

When $\xi_i$ is in $Zone_3(\theta)$ and $\xi_{i-1}$ is in $Zone_1(\theta)$, $\Theta =\emptyset$.
Thus, $f(\theta,x)=0$.

When $\xi_{i-1},\xi_i$ are in $Zone_4(\theta)$, $\Theta =[\xi_{i-1},2\pi-\theta)\cup[\theta,\xi_i+\pi {\rm mod}2\pi)$.
Thus, $f(\theta,x)=r(2\cos\theta-\cos\xi_{i-1}+\cos\xi_i)-x(\pi-2\theta-\xi_{i-1}+\xi_i)$.

When $\xi_i$ is in $Zone_4(\theta)$ and $\xi_{i-1}$ is in $Zone_1(\theta)$, $\Theta =[\theta,\xi_i+\pi {\rm mod}2\pi)$.
Thus, $f(\theta,x)=r(\cos\theta+\cos\xi_i)-x(-\pi-\theta+\xi_i)$.

When $\xi_i$ is in $Zone_4(\theta)$ and $\xi_{i-1}$ is in $Zone_2(\theta)$, $\Theta =[\xi_{i-1},\xi_i+\pi)$.
Thus, $f(\theta,x)=r(\cos\xi_{i-1}+\cos\xi_i)-x(-\xi_{i-1}+\xi_i+\pi)$.


\begin{thebibliography}{99}
\bibitem{smartdust}  [Available online] [accessed on May 26, 2016] http://robotics.eecs.berkeley.edu/\~{ }pister/SmartDust/
\bibitem{wins} G. J. Pottie, et al., Wireless Integrated Network Sensors, Commun. ACM, 43, 5, pp. 51--58, May 2000. 
\bibitem{survey} I. F. Akyildiz, et al.,  A Survey on Sensor Networks, IEEE Communications Magazine, 40, 8, pp. 102--114, 2002.
\bibitem{sensornode} B. W. Cook, et al., SoC Issues for RF Smart Dust, Proceedings of IEEE, 94, 6, pp. 1177--1196, June 2006.
\bibitem{survey2}C. Zhu, et al., A survey on coverage and connectivity issues in wireless sensor networks, J. Network and Computer Applications, 35, pp. 619-632, 2012.
\bibitem{lora} [Available online] [accessed on May 11, 2016] https://www.lora-alliance.org/
\bibitem{nb_iot} [Available online] [accessed on May 9, 2016] http://www.3gpp.org/news-events/3gpp-news/1733-niot 
\bibitem{commag} H. Saito, et al., Wide Area Ubiquitous Network: The Network Operator's View of a Sensor Network, IEEE Communications Magazine, 46, 12, pp. 112-120, 2008.
\bibitem{VTC}M. Umehira, et al., Concept and Feasibility Study of Wide Area Ubiquitous Network for Sensors and Actuators, IEEE VTC 2007 Spring, pp. 165-169, 2007.
\bibitem{sigfox} [Available online] [accessed on May 9, 2016] http://www.sigfox.com/
\bibitem{IoT} J. Gubbi, et al., Internet of Things (IoT): A vision, architectural elements, and future directions, Future Generation Computer Systems, 29, 7, pp. 1645-1660, 2013.
\bibitem{IoTsurvey} L. D. Xu, et al., Internet of Things in Industries: A Survey, IEEE Trans. Industrial Informatics, 10, 4, pp. 2233-2243, 2014. 
\bibitem{infocom} H. Saito, et al., Shape Estimation Using Networked Binary Sensors, INFOCOM 2009.
\bibitem{ieice-invite} H. Saito, Local Information, Observable Parameters, and Global View, IEICE Trans. Communications, E96-B, 12, pp.3017-3027, 2013.
\bibitem{arXiv} H. Saito and H. Honda, Geometric Analysis of Estimability of Target Object Shape Using Location-Unknown Distance Sensors, IEEE Trans. on Control of Network Systems, to appear.

\bibitem{signalProcess} H. Saito, et al., Stochastic Geometric Filter and Its Application to Shape Estimation for Target Objects, IEEE Trans. Signal Processing, 59, 10, , pp. 4971-4984, 2011.
\bibitem{mobileComp} H. Saito, et al., Estimating Parameters of Multiple Heterogeneous Target Objects Using Composite Sensor Nodes, IEEE Trans. Mobile Computing, 11, 1, pp. 125-138, 2012. 

\bibitem{time-variant} H. Saito, et al., Parameter Estimation Method for Time-variant Target Object Using Randomly Deployed Sensors and its Application to Participatory Sensing, IEEE Trans. Mobile Computing, 14, 6, pp. 1259-1271, 2015.
\bibitem{new} Hiroshi Saito and Tatsuaki Kimura, Theoretical Framework for Estimating Target-Object Shape by Using Location-Unknown Mobile Distance Sensors, submitted (arXiv 1802.06882).


\bibitem{locating_nodes} N. Patwari, et al., Locating the nodes, IEEE Signal Processing Magazine, 22, 4, pp. 54-69, 2005.
\bibitem{flip_amb}A. A. Kannan, et al., Analysis of Flip Ambiguities for Robust Sensor Network Localization, IEEE Trans. Vehicular Technology, 59, 4, pp. 2057-2070, 2010.
\bibitem{local_4} G. Mao, et al., Wireless sensor network localization techniques, Computer Networks, 51, 10, pp. 2529-2553, 2007.
\bibitem{local_2}F. Gustafsson, et al., Mobile positioning using wireless networks: Possibilities and fundamental limitations based on available wireless network measurements, IEEE Signal Process. Mag., 22, 4, pp. 41-53, 2005.
\bibitem{local_3} A. Sayed, et al., Network-based wireless location: Challenges faced in developing techniques for accurate wireless location information, IEEE Signal Process. Mag., 22, 4, pp. 24-40, 2005.
\bibitem{tsp2002}J. C. Chen, et al., Maximum-Likelihood Source Localization and Unknown Sensor Location Estimation for Wideband Signals in the Near-Field, IEEE Trans. Signal Processing, 50, 8, pp. 1843-1854, 2002.
\bibitem{acm_sensor} X. Nguyen, et al., A Kernel-Based Learning Approach to Ad Hoc Sensor Network Localization, ACM Trans. Sensor Networks, 1, 1, pp. 134-152, 2005.
\bibitem{bernoulli}B. Ristic, et al., A Tutorial on Bernoulli Filters: Theory, Implementation and Applications, IEEE Trans. Signal Processing, 61, 13, pp. 3406-3430, 2013.

\bibitem{camera}Yibo Wu, Yi Wang, and Guohong Cao, Photo Crowdsourcing for Area Coverage in Resource Constrained Environments, INFOCOM 2017.

\bibitem{mclust} [Available online]  [accessed on January 25, 2016] https://cran.r-project.org/web/packages/mclust/mclust.pdf


\end{thebibliography}
\end{document}